\documentclass[12pt,english,aps,tightenlinesletterpaper,superscriptaddress,secnumarabic,nofootinbib]{revtex4}
\usepackage{mathptmx}

\usepackage[T1]{fontenc}
\usepackage[latin9]{inputenc}
\usepackage{color}
\usepackage{babel}
\usepackage{amsmath}
\usepackage{amssymb}
\usepackage{graphicx}
\usepackage{esint}
\usepackage[unicode=true,pdfusetitle,
 bookmarks=true,bookmarksnumbered=false,bookmarksopen=false,
 breaklinks=false,pdfborder={0 0 1},backref=section,colorlinks=false]
 {hyperref}
\hypersetup{
 citecolor=blue}

\makeatletter
\@ifundefined{textcolor}{}
{%
 \definecolor{BLACK}{gray}{0}
 \definecolor{WHITE}{gray}{1}
 \definecolor{RED}{rgb}{1,0,0}
 \definecolor{GREEN}{rgb}{0,1,0}
 \definecolor{BLUE}{rgb}{0,0,1}
 \definecolor{CYAN}{cmyk}{1,0,0,0}
 \definecolor{MAGENTA}{cmyk}{0,1,0,0}
 \definecolor{YELLOW}{cmyk}{0,0,1,0}
}

\usepackage{amsfonts}\usepackage{bbm}\usepackage{euscript}
\usepackage{dcolumn}
\usepackage{bm}
\usepackage{epsfig}\epsfclipon
\usepackage{slashed}

\makeatother

\begin{document}

\title{Elementary Theorems Regarding Blue Isocurvature Perturbations}

\author{Daniel J. H. Chung}

\email{danielchung@wisc.edu}

\selectlanguage{english}%

\affiliation{Department of Physics, University of Wisconsin, Madison, WI 53706,
USA}

\affiliation{Kavli Institute for Cosmological Physics, University of Chicago,
Chicago IL 60637, USA}

\author{Hojin Yoo}

\email{hyoo@lbl.gov}

\selectlanguage{english}%

\affiliation{Department of Physics, University of Wisconsin, Madison, WI 53706,
USA}

\affiliation{Berkeley Center for Theoretical Physics, University of California,
Berkeley, CA 94720}

\affiliation{Theoretical Physics Group, Lawrence Berkeley National Laboratory,
Berkeley, CA 94720}
\begin{abstract}
Blue CDM-photon isocurvature perturbations are attractive in terms
of observability and may be typical from the perspective of generic
mass relations in supergravity. We present and apply three theorems
useful for blue isocurvature perturbations arising from linear spectator
scalar fields. In the process, we give a more precise formula for
the blue spectrum associated with the axion model of 0904.3800, which
can in a parametric corner give a factor of $O(10)$ correction. We
explain how a conserved current associated with Peccei-Quinn symmetry
plays a crucial role and explicitly plot several example spectra including
the breaks in the spectra. We also resolve a little puzzle arising
from a naive multiplication of isocurvature expression that sheds
light on the gravitational imprint of the adiabatic perturbations
on the fields responsible for blue isocurvature fluctuations.
\end{abstract}
\maketitle
\begin{widetext} \tableofcontents{}\vspace{5mm}
 \end{widetext}

\section{Introduction\label{sec:Introduction}}

Single scalar field inflationary models generate approximately adiabatic,
scale-invariant, and Gaussian primordial density perturbations \cite{Starobinsky:1980te,Sato:1980yn,Linde:1981mu,Mukhanov:1981xt,Albrecht:1982wi,Hawking:1982my,Guth1982,Starobinsky:1982ee,Bardeen:1983qw}.
This is consistent with the Cosmic Microwave Background (CMB) measurements
\cite{Ade:2013xsa,Ade:2013lta,Ade:2013rta,Ade:2013sta,Ade:2013ydc,Hinshaw:2012fq,Komatsu:2010fb,Brown:2009uy,Reichardt:2008ay,Fowler:2010cy,Lueker:2009rx,Hikage:2012be}
and the Large Scale Structure (LSS) observations \cite{Percival:2007yw,Eisenstein:2005su}.
However, non-thermal cold dark matter (CDM) scenarios such as axions
\cite{Peccei:1977hh,Weinberg:1977ma,Wilczek:1977pj} and WIMPZILLAs
\cite{Chung:1998zb,Chung:1998ua,Kuzmin:1999zk,Kuzmin:1998kk,Chung:2001cb,Chung:1998bt}
naturally have observable CDM-photon isocurvature perturbations (e.g.
\cite{Fox:2004kb,Sikivie:2006ni,Preskill:1982cy,Abbott:1982af,Dine:1982ah,Steinhardt:1983ia,Turner:1985si,Hertzberg:2008wr,Kolb:1990vq,Beltran:2006sq,Turner:1983sj,Axenides:1983hj,Linde:1984ti,Seckel:1985tj,Visinelli:2014twa,Choi:2014uaa})
since the CDM never thermalizes with the photons. Indeed, it is remarkable
that a subdominant dark matter component as small as $10^{-4}$ of
the total dark matter content can leave an experimentally detectable
effect through cosmology (see e.g.~\cite{Chung:2011xd}). Furthermore,
isocurvature perturbations are interesting since it can generate rich
density perturbation phenomenology. For example, unlike standard single
field inflationary scenarios, degrees of freedom responsible for isocurvature
perturbations are able to generate large primordial local non-Gaussianities
\cite{Bartolo:2001cw,Chung:2011xd,Kawasaki:2008sn,Langlois:2008vk,Linde:1996gt,Kofman:1989ed,Geyer:2004bx,Ferrer:2004nv,Boubekeur:2005fj,Barbon:2006us,Lyth:2006gd,Koyama:2007if,Lalak:2007vi,Huang:2007hh,Lehners:2008vx,Beltran:2008tc,Kawasaki:2008jy,Langlois:2009jp,Chen:2010xka,Langlois:2010fe,Langlois:2011zz,Mulryne:2011ni,Gong:2011cd,DeSimone:2012gq,Enqvist:2012vx,Kawasaki:2011pd,Langlois:2013dh,Kawasaki:2013ae,Nurmi:2013xv}.

Scale-invariant isocurvature spectrum is well constrained as its power
on CMB length scales has to be less than about 3\% of the adiabatic
power \cite{Crotty:2003fp,Bean:2006qz,Komatsu:2008ex,Valiviita:2009ck,Sollom:2009vd,Komatsu:2010in,Hinshaw:2012fq,Ade:2013rta}.
Because the largest scale invariant isocurvature effects on CMB measurements
occurs on long length scales (e.g.~see the appendix in \cite{Chung:2013sla}),
one expects scale invariant isocurvature effects to be well hidden
in any future observations probing short length scales. However, if
the isocurvature spectrum is very blue, then isocurvature effects
that are hidden on long length scales may become large effects on
short length scales (see e.g.~\cite{Takeuchi:2013hza,Chluba:2013dna,Sekiguchi:2013lma,Dent:2012ne}).
If such strongly blue spectral index isocurvature signal is uncovered
in the future, one may ask what one will learn regarding the high
energy physics of the isocurvature sector. 

One answer to that is given by \cite{Kasuya:2009up} in which a supersymmetric
axion model is constructed giving rise to a blue spectrum. In that
work, the phenomenologically relevant axion isocurvature perturbation
amplitude $\delta a$ is assumed to be given by the frozen value $\delta a/\varphi_{+}$
at horizon crossing where $\varphi_{+}$ is the classical value of
the radial field that breaks the Peccei-Quinn (PQ) symmetry during
inflation. Unlike in the conventional minimal axion scenario in which
the order parameter $\varphi_{+}$ is sitting at its potential minimum
during inflation, its value is initially displaced from the minimum
and is slowly decreasing towards the stable minimum during inflation.
Hence, the assumed frozen value at horizon crossing increases for
larger wave vector $k$ modes which leave the horizon later. In that
way, a blue spectrum is generated over a $k$ range that depends on
the spectral index which controls the amount of time $\varphi_{+}$
takes to settle to its minimum. Furthermore, because supergravity
structure generically induces a Hubble scale mass \cite{Dine:1995uk}
for $\varphi_{+}$, the spectral index can be easily extremely blue.
For example, it has been claimed that this scenario allows an isocurvature
spectral index of $n=4$ for $k\in[k_{\mbox{min}},k_{\mbox{max }}]$
specified by $k_{\mbox{max}}/k_{\mbox{min}}\sim\exp(10).$

Partly motivated by this result, we formulate three elementary ``theorems''
regarding isocurvature perturbations with a very blue spectrum for
non-thermal dark matter fields such as the axions with a displaced
(possibly time dependent) vacuum expectation value during inflation.%
\footnote{The proofs are only at the rigor of a typical physics literature.%
} Theorem 1 defines a superhorizon conserved quantity for systems possessing
an approximate symmetry associated with linearly perturbed system.
The merit of this theorem compared to previous discussions of this
topic in the literature (see e.g.~\cite{Gordon:2000hv,Polarski:1994rz})
is its ability to go beyond the end of inflation and the reheating
process. Theorem 2 describes under what averaging conditions that
fluid quantities behave as $\delta\chi_{nad}/\chi_{0}$. This second
theorem merely restates what is known in the literature in the context
of the theorem. Theorem 3 describes the computation of the quantum
isocurvature perturbations. The merit of theorem 3 compared to the
previous discussion in the literature is the explicit canonical quantization
in the presence of linearized gravitational constraints. Furthermore,
we point out the clear conditions under which the simple analytic
estimates are valid.

We then give couple of applications of our theorems. First, we improve
on the naive quantization of axions in scenario of \cite{Kasuya:2009up}
and compute $O(1+\frac{1}{(n-4)^{2}})$ corrections to the spectrum
of with spectral index $n$. In the process, an interesting application
of conserved PQ symmetry current is made, which explains how two independent
dynamical degrees of freedom behave as a single one during a finite
time duration of interest. The theorems also help to set a precise
boundary of where the simple analytic computations are invalid. For
example, contrary to claims of \cite{Kasuya:2009up}, $n=4$ spectrum
cannot be generated in their scenario. Another consequence of understanding
the boundary is that if the ratio of axion isocurvature blue power
amplitude to the adiabatic power amplitude is at most of the order
of a few percent on the largest observable scales and can be described
in terms of quantization methods presented in this paper (and implicitly
approximated in \cite{Kasuya:2009up}), most of the cold dark matter
must be made of different species. We also illustrate through example
plots, phenomenologically interesting parametric corners of the model
(having a six dimensional parameter space). Although observable spectra
can contain breaks, these break regions typically contain $k$-space
domains for which the simple analytic computation is invalid. We identify
how large the expansion rate $H$ during inflation can be in this
class of models generating a large blue spectrum. A measurement of
tensor-to-scalar ratio at the level of $r=O(10^{-1})$ will disfavor
this class of models, at least in its simplest form. 

In another application, theorem 3 is used to explain why the isocurvature
blue spectrum does not have a simple lower bound suggested by a naive
operator product analysis. More explicitly, the isocurvature perturbations
are defined to be a contrast of the form $S_{\chi}\sim C_{1}\delta\chi-C_{2}\delta\phi$
where $C_{i}$ are background field dependent coefficients and $\delta\phi$
is the inflaton field and $\delta\chi$ is the field responsible for
the existence of isocurvature perturbations. In other words, the isocurvature
field is always dressed with the inflaton sector. The quantum correlator
$\langle SS\rangle$ would then naively have a piece that is proportional
to $C_{2}^{2}\langle\delta\phi\delta\phi\rangle$ coming from the
dressing. This piece for a blue spectrum is of order of the adiabatic
perturbation power spectrum. If the cross correlation piece does not
precisely cancel this piece, $\langle S_{\chi}S_{\chi}\rangle$ would
be of the order of adiabatic spectrum, leading to a simple lower bound.
However, the theorem shows that the power spectrum of $S_{\chi}$
generically behaves independently of the adiabatic spectrum. The more
broad lesson encapsulated by theorem 3 is that the gravitational coupling
of $\delta\chi$ to $\delta\phi$ makes $\delta\chi$ grow an inhomogeneity
that looks like $\delta\phi$ such that $S_{\chi}$ becomes independent
of $\delta\phi$.

The order of the presentation will be as follows. In Sec\@.~\ref{sec:Useful-Simple-Theorems},
three simple theorems and couple of corollaries useful for spectator
dark matter isocurvature spectra are presented. In Sec.~\ref{sec:Applications},
a couple of applications of the theorems are given. One application
corresponds to improving and elucidating the computation of \cite{Kasuya:2009up}.
The second application corresponds to understanding how dressing effects
coming from the definition of the isocurvature perturbations do not
mix inflaton field quantum fluctuations with the dark matter field
quantum fluctuations because of the secular growth imprinting an adiabatic
inhomogeneity to the dark matter field. We close with a summary and
thoughts on future work to be done in this direction. In the appendix,
we collect some results useful for the theorems.

\section{\label{sec:Useful-Simple-Theorems}Useful Simple Theorems for Blue
Isocurvature Models with a Slowly Rolling Time Dependent VEV}

\subsection{Definitions}

In this subsection, we define the language used for our theorems.

\paragraph{Metric and Fourier Conventions}

Although theorems that we present are gauge invariant, we will have
the occasion to use several gauges in our proofs. The Newtonian gauge
scalar perturbations will be parameterized as
\begin{equation}
ds^{2}=(1+2\Psi^{(N)})dt^{2}-a^{2}(t)(1+2\Phi^{(N)})|d\vec{x}|^{2}.\label{eq:metric_Newtonian}
\end{equation}
We consider slow-roll inflaton field $\varphi$ scenarios in which
superhorizon adiabatic perturbations are approximately conserved.
Conserved adiabatic curvature perturbations on superhorizon scales
is given in Newtonian gauge by the solution \cite{Weinberg:2003sw,Wands:2000dp,Mukhanov:1990me,Bardeen:1983qw}
\begin{equation}
\Phi^{(N)}(t,\vec{k})-H\frac{\delta\rho^{(N)}(t,\vec{k})}{\dot{\rho}(t)}=\zeta_{\vec{k}}\equiv\mbox{constant}\label{eq:conserved}
\end{equation}
where $H\equiv\dot{a}/a$ and $|\vec{k}/a|\ll H$ and we have introduced
the Fourier convention
\begin{equation}
Q(t,\vec{k})=\int d^{3}xe^{-i\vec{k}\cdot\vec{x}}Q(t,\vec{x}).\label{eq:Fourierconvention}
\end{equation}
In the Newtonian gauge, the expansion is manifestly isotropic. Furthermore,
$\Phi^{(N)}$ has the intuitive interpretation of being the gravitational
potential in the Poisson equation. Because of these properties, the
field equations in the Newtonian gauge are convenient to work with
when working with classical equations. 

On the contrary, the spatially flat gauge is more useful for quantization
during inflation (see e.g.~\cite{Hwang:1996gz}). The scalar metric
perturbation convention in spatially flat gauge can be chosen to be
\begin{equation}
ds^{2}=(1+2\Psi^{(sf)})dt^{2}+a\partial_{i}F^{(sf)}dtdx^{i}-a^{2}(t)|d\vec{x}|^{2}.
\end{equation}
As shown in Sec. \ref{sub:Theorem3}, the relevant interaction action
derived from solving the gravitational constraints in this gauge consist
only of local terms of the fields unlike for the corresponding equations
in the Newtonian gauge. Thus, the quantization of fields and the investigation
of the subhorizon mode functions are technically simpler in this gauge.
Hence, we will employ the spatially flat gauge only for quantization
during inflation which establishes the initial conditions for the
late time classical equations.

\paragraph{Linear Spectator Isocurvature Field}

Let linear spectator isocurvature field be defined as a canonically
normalized scalar field $\chi=\chi_{0}(t)+\delta\chi^{(N)}(t,\vec{x})$
for which
\begin{equation}
\delta\rho_{\chi}^{(N)}\propto\delta\chi^{(N)}+O(\delta\chi^{(N)\,2})
\end{equation}
\begin{equation}
\chi_{0}^{(N)}(t_{\mbox{during inflation}})\gg\frac{H}{2\pi}|_{\mbox{during inflation}}
\end{equation}
\begin{equation}
\frac{\delta\rho_{\chi}^{(N)}}{\delta\rho_{\mbox{dominant}}^{(N)}}=\frac{\delta T_{\chi\,\,\,\,\,\,\,0}^{(N)\,0}}{\delta T_{\mbox{dominant}\,\,\,\,\,\,\,0}^{(N)\,0}}\ll1\label{eq:smallenergyassumption}
\end{equation}
in Newtonian gauge where the subscript ``dominant'' corresponds
to the energy density component that dominates $T_{\,\,\,\,\,\,\,\,\,0}^{(N)\,0}$.
For example, during inflation, ``dominant'' corresponds to the label
$\varphi$ while during radiation domination, ``dominant'' corresponds
to the label $\gamma$ representing the relativistic degrees of freedom.
Because we will focus on 
\begin{equation}
V_{\chi}=\frac{1}{2}m^{2}\chi^{2},
\end{equation}
Eq.~(\ref{eq:smallenergyassumption}) translates to
\begin{equation}
\frac{m^{2}}{H^{2}}\frac{\chi_{0}}{M_{p}}\ll3\sqrt{2\epsilon}\label{eq:expansioninfield}
\end{equation}
where $\epsilon$ is the inflationary slow-roll parameter. If the
inflaton potential is given as $V_{\varphi}(\varphi)$, then
\begin{equation}
\epsilon=\frac{M_{p}^{2}}{2}\left(\frac{V_{\varphi}'(\varphi_{0})}{V_{\varphi}(\varphi_{0})}\right)^{2}
\end{equation}
where $\varphi=\varphi_{0}(t)+\delta\varphi$. The effective expansion
parameters are $\chi_{0}/M_{p}$ and slow-roll parameters of the inflaton
field if $m/H\sim O(1)$.

\paragraph{Spectral Conventions}

The gauge invariant spectrum of linear spectator $\chi$-photon isocurvature
perturbations useful for Boltzmann equations is often defined \emph{during
radiation dominated universe} through 
\begin{equation}
\Delta_{s_{\chi}}^{2}(k)\equiv\frac{k^{3}}{2\pi^{2}}\int\frac{d^{3}k'}{(2\pi)^{3}}\langle\delta_{s_{\chi}}(t,\vec{k})\delta_{s_{\chi}}(t,\vec{k}')\rangle\label{eq:spectrum-def}
\end{equation}
\begin{equation}
\delta_{s_{\chi}}\equiv3(\zeta_{\chi}-\zeta_{\gamma})\label{eq:deltas}
\end{equation}
\begin{equation}
\zeta_{\chi}=\Phi^{(N)}+\frac{\delta\rho_{\chi}^{(N)}|_{\mbox{background smoothed}}}{3\langle\rho_{\chi}+P_{\chi}\rangle_{\mbox{time}}},\,\,\,\,\,\zeta_{\gamma}=\Phi^{(N)}+\frac{\delta\rho_{\gamma}^{(N)}}{3(\rho_{\gamma}+P_{\gamma})}\label{eq:zetadefsRD}
\end{equation}
where $P_{i}$ are pressure quantities corresponding to $-T_{\,\,\, i}^{i}$
components of the energy momentum tensor. Here, the ``time'' average
in the denominator of the definition of $\zeta_{\chi}$ corresponds
to a time average over $m^{-1}$ time scale. The ``background smoothed''
in the numerator of the definition of $\zeta_{\chi}$ corresponds
to averaging over $m^{-1}$ time scale all quadratic terms in the
background $\chi_{0}(t)$ appearing in the numerator. The variables
$\zeta_{\chi}$ and $\zeta_{\gamma}$ are conserved outside the horizon
if the pressure of the constituent is a function only of its energy
density. In particular, $\zeta_{\gamma}$ corresponds to the gauge-invariant
curvature perturbation if we assume that radiation behaves as a single
component fluid coming from the inflaton decay. For single-field inflation,
observational normalization of 
\begin{eqnarray}
\Delta_{\zeta}^{2}(k_{0}=0.05\mbox{Mpc}^{-1}) & = & \Delta_{\zeta_{\gamma}}^{2}(k_{0})\nonumber \\
 & \approx & \frac{V_{\varphi}(k_{0})}{24\pi^{2}M_{p}^{4}\epsilon}\approx2.4\times10^{-9}\label{eq:slow-rollresult}
\end{eqnarray}
 corresponds to the currently known approximate value of adiabatic
curvature perturbation amplitude.

The $\chi$-photon isocurvature spectrum often contain $k$-space
domains which can be parameterized as
\begin{equation}
\Delta_{s_{\chi}}^{2}(k)=\Delta_{s_{\chi}}^{2}(k_{0})\left(\frac{k}{k_{0}}\right)^{n-1}
\end{equation}
where $n$ is the spectral index. The isocurvature spectrum is blue
when $n>1$. The primary focus of this paper is regarding spectra
for which $n-1\gtrsim O(0.1)$ which become parametrically insensitive
to the inflationary slow roll parameter values of $O(\epsilon)<0.02$.
As far as the phenomenological bounds are concerned, note that
\begin{equation}
\Delta_{s}^{2}(k)=\omega_{\chi}^{2}\Delta_{s_{\chi}}^{2}(k)\label{eq:mixturefirst}
\end{equation}
where $\omega_{\chi}\leq1$ is the fraction of cold dark matter that
is in the $\chi$ field as is explained in Appendix \ref{sec:Mixture}.
The current phenomenological bounds on $\Delta_{s}^{2}(k)/\Delta_{s_{\chi}}^{2}(k)$
for scale invariant power spectrum is approximately a few percent
\cite{Crotty:2003fp,Bean:2006qz,Komatsu:2008ex,Valiviita:2009ck,Sollom:2009vd,Komatsu:2010in,Hinshaw:2012fq,Ade:2013rta}.

With these definitions and assumptions, we can construct a useful
statement that can be used to set classical equation boundary conditions
before Eq.~(\ref{eq:smallenergyassumption}) breaks down. The most
important of the three theorems that will be presented below is theorem
three. Note that one of the key merits of the theorem that we are
presenting is its applicability connecting computations during inflation
to variables during radiation domination.

\subsection{Theorem 1: Classically Conserved Isocurvature Quantity}

Here is a statement of the first theorem. In slow-roll inflationary
scenarios, the linear spectator isocurvature quantity
\begin{equation}
S_{\chi}(t,\vec{k})\equiv\frac{2\delta\chi_{nad}}{\chi_{0}(t)}\label{eq:isocurvaturequantity}
\end{equation}
where
\begin{equation}
\delta\chi_{nad}\equiv\delta\chi^{(G)}(t,\vec{k})-\delta\chi_{ad}^{(G)}(t,\vec{k})
\end{equation}
on superhorizon length scales is approximately conserved as long as
$\chi$ interaction is dominated by $V_{\chi}=m^{2}\chi^{2}/2$ \emph{and
gravity}, anisotropic stress effects can be neglected, and attractor
behavior of $\delta\chi_{nad}$ and $\chi_{0}(t)$ is relevant \emph{during
inflation} (i.e. non-pathological boundary conditions are chosen for
the homogeneous field) with an expansion rate of $H$. A sufficient
condition for attractor behavior with non-pathological boundary condition
is
\[
|\dot{\chi}_{0}(t_{\mbox{iniital}})|\lesssim m^{2}\chi_{0}(t_{\mbox{initial}})/H
\]
\begin{equation}
\nu N_{k}\gg1
\end{equation}
 where 
\begin{equation}
\nu\equiv\frac{3}{2}\sqrt{1-\frac{4}{9}\frac{m^{2}}{H^{2}}}
\end{equation}
and $N_{k}$ is the number of efolds between the time of $k$-mode
horizon exit and the end of inflation. Here, we have defined
\begin{equation}
\delta\chi_{ad}^{(G)}(t,\vec{k})\equiv-\zeta_{\vec{k}}\frac{\dot{\chi}_{0}(t)}{a(t)}\int dta(t)+\xi^{0}\partial_{0}\chi_{0}(t)
\end{equation}
where $\xi^{0}=0$ in the Newtonian gauge (i.e. $G=N$) and in any
other gauge $G$ is related to the Newtonian gauge coordinates through
$x^{(N)\,\mu}=x^{(G)\,\mu}+(\xi^{0},\delta^{ij}\partial_{i}\xi)$.
Furthermore, $S_{\chi}$ is a \emph{gauge invariant} quantity. Note
that we have introduced a factor of 2 in the definition of $S_{\chi}$
for later convenience. Finally, note that this theorem is formulated
at the classical solution level. The error in the conservation coming
from the assumption of attractor behavior can be estimated as
\begin{equation}
\mbox{attractor error }\sim O\left(\exp\left[-2\nu N_{k}\right]\right)
\end{equation}
where the coefficient of the error depends on details of initial conditions
of both the homogeneous mode and the perturbation mode at the beginning
of inflation. The fractional error $O(\mathcal{E})$ in the conservation
is approximated to be the terms that are dropped in making this statement:
\begin{equation}
\mathcal{E}=\exp\left[-2\nu N_{k}\right]+\frac{\delta\rho_{\chi}^{(N)}}{\delta\rho_{\mbox{dominant}}^{(N)}}.
\end{equation}
We also implicitly assume that the post-inflationary cosmological
history consists of smoothly connected patches of power-laws.

It is important to note that this theorem makes $S_{\chi}$ conserved
independently of the details not stated in the theorem, including
some of the details of the end of the inflation, reheating, early
radiation domination, and how $\chi_{0}$ makes the transition from
a slow-roll field to a coherently oscillating one. In particular,
the classical conservation here is valid even when $\epsilon\rightarrow1$
at the end of inflation, unlike the spatially flat gauge quantity
$\delta\chi^{(sf)}/\chi_{0}$ which undergoes generically undergoes
time evolution at the end of inflation. The conditions stated in the
theorem can be understood as a decoupling limit of the isocurvature
perturbations, and this theorem establishes a classically conserved
quantity in that limit. Note one of the important points for this
paper: the numerator $\delta\chi$ and the denominator $\chi_{0}$
must correspond to the same dynamical degree of freedom that responds
to the same potential $V_{\chi}$ dominated by the mass term. Finally,
note that when we state the assumption that the mass term and gravity
dominate the interactions, we are stating that perturbative interactions
are too weak to thermalize the system.
\begin{description}
\item [{proof}]~
\end{description}
Consider the equation of motion for the perturbation variable in the
Newtonian gauge $\delta\chi^{(N)}$ in the long wavelength limit in
which we can neglect the gradient terms:
\begin{equation}
\delta\ddot{\chi}^{(N)}+3H\delta\dot{\chi}^{(N)}+V_{\chi}''(\chi_{0})\delta\chi^{(N)}-4\dot{\chi}_{0}\dot{\Psi}^{(N)}+2V_{\chi}'(\chi_{0})\Psi^{(N)}=0.\label{eq:eom}
\end{equation}
Here, we have assumed that gravitational interactions and potential
self-interactions $V_{\chi}'(\chi_{0})$ dominate the interactions.
If anisotropic stress effects can be neglected, the $ij$ component
of Einstein equations imply
\begin{equation}
\Phi^{(N)}=-\Psi^{(N)}.
\end{equation}
The 00 component of Einstein equation in Newtonian gauge partially
determining $\Psi^{(N)}$ is
\begin{equation}
-3\frac{\dot{a}}{a}(H\Psi^{(N)}+\dot{\Psi}^{(N)})=\frac{1}{2M_{p}^{2}}\left[\delta\rho_{\chi}^{(N)}+\delta\rho_{\mbox{dominant}}^{(N)}\right]
\end{equation}
where
\begin{equation}
\delta\rho_{\chi}^{(N)}=-\dot{\chi}_{0}^{2}\Psi+\dot{\chi}_{0}\partial_{t}\delta\chi+V_{\chi}'\delta\chi
\end{equation}
and $\delta\rho_{\mbox{dominant}}^{(N)}\equiv\delta T_{\mbox{dominant}\,\,\,\,0}^{0\,(N)}$
is the dominant contribution to the energy-momentum tensor as discussed
in Eq.~(\ref{eq:smallenergyassumption}).%
\footnote{These statements can easily be covariantized, but such formalizations
tend to obscure the intuition rather than to illuminate the intuition.
Since our aim is to illuminate the intuition of the simple physics,
we will leave the presentation in the explicitly gauge dependent form.%
} In the limit
\begin{equation}
\frac{\delta\rho_{\chi}^{(N)}}{\delta\rho_{\mbox{dominant}}^{(N)}}\ll1,\label{eq:condition}
\end{equation}
we see that $\Psi^{(N)}$ is independently of $\delta\chi$. Hence,
with the condition of Eq.~(\ref{eq:condition}), the $\Psi^{(N)}$
dependent terms in Eq.~(\ref{eq:eom}) are external sources terms.
Due to dilatation diffeomorphism gauge solution that lifts to physical
solutions \cite{Weinberg:2004kr}, there exists an adiabatic solution
\begin{equation}
\delta\chi_{ad}^{(N)}=-\zeta_{\vec{k}}\frac{\dot{\chi}_{0}(t)}{a(t)}\int dta(t)\label{eq:particular}
\end{equation}
where $\zeta_{\vec{k}}$ is the usual time independent gauge-invariant
curvature perturbation constant determined by the inflaton sector
$\varphi$ approximately independently of $\delta\chi^{(N)}$ as long
as Eq.~(\ref{eq:condition}) is satisfied. From the perspective of
the classical equations we are discussing here, $\zeta_{k}$ is simply
a constant parameterizing a solution to Eq.~(\ref{eq:eom}) where
the gravitational potential is given by Eq.~(\ref{eq:conserved}).%
\footnote{Although we have not made any explicit assumptions about the background
energy density, Eq.~(\ref{eq:condition}) does depend on the background
energy density.%
}

Given that Eq.~(\ref{eq:eom}) is a second order differential equation,
the most general solution corresponds to two independent solutions
$h_{1,2}$ to the homogeneous equation added to the particular solution
given by Eq.~(\ref{eq:particular}):
\begin{equation}
\delta\chi^{(N)}=c_{1}h_{1}^{(N)}+c_{2}h_{2}^{(N)}+\delta\chi_{ad}^{(N)}
\end{equation}
where $c_{1}$ and $c_{2}$ are coefficients independent of time.
Hence, we see that in the limit that $k/(aH)\rightarrow0$ can be
neglected, the numerator of 
\begin{equation}
\frac{\delta\chi_{nad}}{\chi_{0}(t)}=\frac{\delta\chi^{(N)}-\delta\chi_{ad}^{(N)}}{\chi_{0}(t)}=\frac{c_{1}(\vec{k})h_{1}^{(N)}(t,\vec{k})+c_{2}(\vec{k})h_{2}^{(N)}(t,\vec{k})}{\chi_{0}(t)}
\end{equation}
is governed by the same equation as the denominator if $V_{\chi}=m^{2}\chi^{2}/2$:
i.e.
\begin{equation}
\ddot{h}_{i}^{(N)}+3H\dot{h}_{i}^{(N)}+m^{2}h_{i}^{(N)}=0
\end{equation}
\begin{equation}
\ddot{\chi}_{0}^{(N)}+3H\dot{\chi}_{0}^{(N)}+m^{2}\chi_{0}^{(N)}=0.
\end{equation}
Hence, we can also write 
\begin{equation}
\chi_{0}=e_{1}h_{1}^{(N)}+e_{2}h_{2}^{(N)}.
\end{equation}

We know that one mode decays faster than the other during inflation.
This is what we usually call the attractor behavior during inflation
\cite{Mukhanov:1990me}. We will call the less decaying mode $h_{1}^{(N)}$.
More quantitatively, in the dS approximation, we have
\begin{equation}
\left|\frac{h_{1}^{(N)}}{h_{2}^{(N)}}\right|=e^{2H\nu t}.
\end{equation}
 In this case, we thus have at the end of inflation (when this relative
growth ends)
\begin{equation}
\frac{\delta\chi^{(N)}-\delta\chi_{ad}^{(N)}}{\chi_{0}(t)}=\frac{c_{1}(\vec{k})+c_{2}(\vec{k})O(e^{-2\nu N_{k}})}{e_{1}+e_{2}(\vec{k})O(e^{-2\nu N_{k}})}
\end{equation}
which is independent of time $N_{k}$ (the number of scale factor
efolds between $k$ mode horizon exit and ) as long as
\begin{equation}
\nu N_{k}\gg1.
\end{equation}
Hence, the error in the conservation coming from the attractor assumption
is $O\left(\exp\left[-2\nu N_{k}\right]\right).$

Finally, under the gauge transformation $x^{(N)\,\mu}=x^{(G)\,\mu}+(\xi^{0},\delta^{ij}\partial_{i}\xi)$
we have
\begin{eqnarray}
\delta\chi^{(G)}(t,\vec{k}) & = & \delta\chi^{(N)}(t,\vec{k})+\xi^{0}\partial_{0}\chi_{0}(t)\\
\delta\chi_{ad}^{(G)}(t,\vec{k}) & = & \delta\chi_{ad}^{(N)}(t,\vec{k})+\xi^{0}\partial_{0}\chi_{0}(t)
\end{eqnarray}
on long wavelengths. This means
\begin{equation}
\frac{\delta\chi^{(N)}-\delta\chi_{ad}^{(N)}}{\chi_{0}(t)}=\frac{\delta\chi^{(G)}-\delta\chi_{ad}^{(G)}}{\chi_{0}(t)}
\end{equation}
for general gauges $G$ non-singularly connected to the Newtonian
gauge $N$. We thus see that this quantity is gauge invariant.

Note that one may wonder whether there are other interactions besides
mass interactions that would lead to the same result. To see that
this is not generically possible with only potential modifications,
note that for $\delta\chi^{(N)}-\delta\chi_{ad}^{(N)}$ to behave
similarly as $\chi_{0}(t)$, a generic condition is 
\begin{equation}
V_{\chi}'(\delta\chi)\approx V_{\chi}''(\delta\chi)\delta\chi
\end{equation}
which can easily be solved to obtain
\begin{equation}
V_{\chi}(\delta\chi)\approx C_{1}\delta\chi^{2}+C_{2}
\end{equation}
which means that $\chi$ interaction is dominated by the mass term.

A trivial corollary of this theorem is to discuss the situation when
the constant $m$ is replaced by $m(t)$ which is constant during
a finite time interval during inflation and makes a transition to
another value during inflation. 
\begin{description}
\item [{corollary~1}] In the context of theorem 1, suppose $m$ is not
a constant but makes a transition to another value during inflation:
\begin{equation}
m^{2}(t)=\begin{cases}
m_{1} & \,\,\,\,\,\, t<t_{c}\\
m_{2} & \,\,\,\,\,\, t>t_{c}
\end{cases}
\end{equation}
where the transition time region near $t=t_{c}$ is assumed to be
much smaller in time than $m_{1}^{-1}$ and $H^{-1}$. The quantity
$S_{\chi}$ is still conserved as long as sufficient time has passed
during the $t<t_{c}$ period to be in the attractor approximation
just as in theorem 1: i.e.
\begin{equation}
H(t_{c}-t_{k})\nu(m_{1})\gg1
\end{equation}
where 
\begin{equation}
\nu(m_{1})=\frac{3}{2}\sqrt{1-\frac{4}{9}\frac{m_{1}^{2}}{H^{2}}}.
\end{equation}
The error estimate associated with this conservation is $O(\mathcal{E})$
where
\begin{equation}
\mathcal{E}\equiv\max\left\{ \exp\left[-2\nu(m_{1})(t_{c}-t_{k})H\right],\,\,\frac{\delta\rho_{\chi}^{(N)}}{\delta\rho_{\mbox{dominant}}^{(N)}}\right\} 
\end{equation}
again with the neglect of any possible secular effects that depend
on unusual cosmological histories.
\item [{proof}]~
\end{description}
On superhorizon scales, we have just as in theorem 1 proof
\begin{equation}
\ddot{h}_{i}^{(N)}+3H\dot{h}_{i}^{(N)}+m^{2}(t)h_{i}^{(N)}=0
\end{equation}
\begin{equation}
\ddot{\chi}_{0}^{(N)}+3H\dot{\chi}_{0}^{(N)}+m^{2}(t)\chi_{0}^{(N)}=0.
\end{equation}
Attractor behavior during $t<t_{c}$ gives for the solutions 
\begin{eqnarray}
\chi_{0} & \approx & e_{1}h_{1}^{(N)}(1+O(e^{-2\nu(m_{1})H(t_{c}-t_{k})}))\\
\delta\chi_{nad} & \approx & c_{1}h_{1}^{(N)}(1+O(e^{-2\nu(m_{1})H(t_{c}-t_{k})}))
\end{eqnarray}
in the language of the proof of theorem 1. Using the well known ``sudden''
approximation, one can match these solutions valid for $t<t_{c}$
to those valid for $t>t_{c}$:
\begin{eqnarray}
e_{1}\left(\begin{array}{c}
h_{1}^{(N)}\\
\dot{h}_{1}^{(N)}
\end{array}\right)_{t=t_{c}} & = & \left(\begin{array}{cc}
H_{1}^{(N)} & H_{2}^{(N)}\\
\dot{H}_{1}^{(N)} & \dot{H}_{2}^{(N)}
\end{array}\right)_{t=t_{c}}\left(\begin{array}{c}
E_{1}\\
E_{2}
\end{array}\right)\\
c_{1}\left(\begin{array}{c}
h_{1}^{(N)}\\
\dot{h}_{1}^{(N)}
\end{array}\right)_{t=t_{c}} & = & \left(\begin{array}{cc}
H_{1}^{(N)} & H_{2}^{(N)}\\
\dot{H}_{1}^{(N)} & \dot{H}_{2}^{(N)}
\end{array}\right)_{t=t_{c}}\left(\begin{array}{c}
C_{1}\\
C_{2}
\end{array}\right)
\end{eqnarray}
where $E_{i}$ and $C_{i}$ are independent solution coefficients
specifying the $\chi_{0}$ and $\delta\chi_{nad}$ solutions (respectively)
in the time region $t>t_{c}$ and $H_{i}^{(N)}$ are independent solutions
in the $t>t_{c}$ time region. Clearly, we have in the region $t>t_{c}$
\begin{equation}
\frac{\delta\chi_{nad}}{\chi_{0}}=\frac{c_{1}}{e_{1}}
\end{equation}
which is a constant.

\subsection{Theorem 2: Gauge Invariant Isocurvature Spectrum During Radiation
Domination}

Here is a statement of theorem 2. The radiation dominated period linear
spectator isocurvature perturbation spectrum defined by Eq.~(\ref{eq:spectrum-def})
on superhorizon length scales is given by
\begin{equation}
\Delta_{s_{\chi}}^{2}(k)=\frac{k^{3}}{2\pi^{2}}\int\frac{d^{3}k'}{(2\pi)^{3}}\langle S_{\chi}(t,\vec{k})S_{\chi}(t,\vec{k}')\rangle\label{eq:result}
\end{equation}
which is time independent as long as $\chi$ interaction is dominated
by $V_{\chi}=m^{2}\chi^{2}/2$ and gravity, anisotropic stress effects
can be neglected, slow-roll attractor behavior of the $\chi_{0}(t)$
\emph{during inflation} is relevant (e.g. boundary conditions close
to slow-roll are chosen for the homogeneous field), and $m\gg3H/2$
during the radiation dominated time period when one wishes to evaluate
this expression. An important part of the linear spectator requirement
is given by Eq.~(\ref{eq:smallenergyassumption}). An interesting
point of this theorem is that $\Delta_{s_{\chi}}^{2}$ is defined
in Eq.~(\ref{eq:spectrum-def}) with $\dot{\chi}_{0}^{2}$ in the
denominator while Eq.~(\ref{eq:result}) is proportional to $\chi_{0}$
in the denominator. 
\begin{description}
\item [{Proof}]~
\end{description}
Consider the computation of
\begin{equation}
\Delta_{s_{\chi}}^{2}(k)=\frac{k^{3}}{2\pi^{2}}\int\frac{d^{3}k'}{(2\pi)^{3}}\langle\delta_{s_{\chi}}(t,\vec{k})\delta_{s_{\chi}}(t,\vec{k}')\rangle.
\end{equation}
during radiation domination when $m\gg3H/2$. Since $\chi_{0}(t)$
is coherently oscillating, its energy density consists mostly of non-relativistic
energy density. Hence, the definition of 
\begin{equation}
\delta\chi_{nad}\equiv\delta\chi^{(N)}-\delta\chi_{ad}^{(N)}\label{eq:nonad}
\end{equation}
gives
\begin{equation}
\delta_{s_{\chi}}=\frac{\dot{\chi}_{0}\partial_{t}\delta\chi_{nad}|_{\mbox{background smoothed}}+m^{2}\delta\chi_{nad}|_{\mbox{background smoothed}}}{\langle\dot{\chi}_{0}^{2}\rangle_{\mbox{time}}}
\end{equation}
according to Eq.~(\ref{eq:zetadefsRD}) where we also defined the
term ``background smoothed.''

From the definition of the conserved quantity $S_{\chi}^{(G)}$ in
Eq.~(\ref{eq:isocurvaturequantity}), we can make the substitution
\begin{equation}
\delta\chi_{nad}=\frac{\chi_{0}(t)}{2}S_{\chi}.
\end{equation}
Hence, the isocurvature classical quantity during radiation domination
simplifies to
\begin{equation}
\delta_{s_{\chi}}=\frac{S_{\chi}}{2}\left[1+\frac{m^{2}\langle\chi_{0}^{2}\rangle_{\mbox{time}}}{\langle\dot{\chi}_{0}^{2}\rangle_{\mbox{time}}}\right]
\end{equation}
where we have used theorem 1 in keeping $S_{\chi}$ constant, and
the time average is given by
\begin{eqnarray}
\langle\rho+P\rangle_{\mbox{time}} & = & \langle\dot{\chi}_{0}^{2}\rangle_{\mbox{time}}\\
 & = & m^{2}\langle\chi_{0}^{2}\rangle_{\mbox{time}}.
\end{eqnarray}
This gives
\begin{equation}
\delta_{s_{\chi}}=S_{\chi}.
\end{equation}
We thus conclude that $\Delta_{s_{\chi}}^{2}(k)$ is asymptotically
time invariant during this radiation dominated time period when the
conditions of theorem 1 are valid approximations.

\subsection{Theorem 3: Quantum Correlator of Linear Spectator Isocurvature Perturbations\label{sub:Theorem3}}

In addition to the conditions of theorem 2, if Bunch-Davies boundary
conditions to the inflationary quantization are imposed, $m<3H(\mbox{during inflation)}/2$,
and a slow-roll inflationary phase characterized by the slow-roll
function $\epsilon\ll1$ defined by $\dot{H}=-\epsilon H^{2}$ during
inflation occurs, the spectrum of linear spectator isocurvature perturbations
on superhorizon scales during radiation domination is given by
\begin{equation}
\Delta_{s_{\chi}}^{2}(k)\approx4\left(\frac{2^{2\nu-1}|\Gamma(\nu)|^{2}}{\pi}\right)\left(\frac{H(t_{k_{0}})}{2\pi\chi_{0}(t_{k_{0}})}\right)^{2}\left(\frac{k}{k_{0}}\right)^{3-2\nu-2\epsilon_{k_{0}}+O(\epsilon_{k_{0}})g_{0}(m/H)}\label{eq:mainfinresult}
\end{equation}
where $t_{k}$ satisfies $k/a(t_{k})=H(t_{k})$ during inflation,
$g_{0}(m/H)$ is an order unity function which vanishes when $m/H\rightarrow0$,
and
\begin{equation}
\nu\equiv\frac{3}{2}\sqrt{1-\frac{4}{9}\frac{m^{2}}{H^{2}(t_{k_{0}})}}\label{eq:nudefintheorem}
\end{equation}
with $k_{0}$ being a fiducial wave vector (typically taken on CMB
length scales). Here, this expression is only valid when 
\begin{equation}
\nu\tilde{N}_{k}\gg|\ln f|\label{eq:boundonnu}
\end{equation}
where $\tilde{N}_{k}\in\{f/\epsilon_{k},\, H(t_{e}-t_{k})\}$. Here,
$t_{k}$ is the time of the horizon exit of mode $k$, $t_{e}$ is
the end of inflation, and $f$ is the fractional error tolerance in
the computation. Moreover, for $\nu$ to be real and also for the
correction to $2^{\nu}\left|\Gamma(\nu)\right|^{2}$ to be within
the error tolerance throughout the time period between when the mode
$k>k_{0}$ leaves the horizon and the mode $k_{0}$ leaves the horizon,
we must have
\begin{eqnarray}
2\varepsilon_{k_{0}}\frac{m^{2}}{H^{2}(t_{k_{0}})\nu^{2}}\ln\left(\frac{k}{k_{0}}\right) & <f & \,\,\,\,\,\,\mbox{for }k>k_{0}.\label{eq:constr_nu}
\end{eqnarray}
The bounds on $\nu$ sets the largest valid isocurvature spectral
index of 
\begin{equation}
n=4-2\nu\label{eq:isoindex}
\end{equation}
coming from Eq.~(\ref{eq:mainfinresult}). Finally, since $\chi_{0}(t_{k})$
was approximated by expanding about $t_{k_{0}},$ we must have
\begin{equation}
k_{0}\exp\left(-\frac{f}{\epsilon_{k_{0}}}\right)<k<k_{0}\exp\left(\frac{f}{\epsilon_{k_{0}}}\right).\label{eq:connectedthroughdS}
\end{equation}
The fractional error $O(\mathcal{E})$ in Eq.~(\ref{eq:mainfinresult})
is approximated to be the terms that are dropped in making this statement:
\begin{equation}
\mathcal{E}=\exp\left[-2\nu(m_{1})\tilde{N}_{k}\right]+\frac{\delta\rho_{\chi}^{(N)}}{\delta\rho_{\mbox{dominant}}^{(N)}}+\frac{m^{2}}{3\sqrt{2\epsilon}H^{2}}\frac{|\chi_{0}(t_{k})|}{M_{p}}+\frac{|\chi_{0}(t_{k})|}{M_{p}}+\epsilon_{k_{0}}+f.\label{eq:errorestimate}
\end{equation}
Because we are neglecting $O(\epsilon_{k_{0}})$ contributions, this
result is applicable to very blue spectra in which $3-2\nu\gg\epsilon_{k_{0}}$,
and because $g_{0}(0)=0$, this result is applicable to $3-2\nu=0$
case. 

Before moving onto the proof, let us make some comments. One immediate
implication of this theorem is that large isocurvature spectra with
blue spectral indices are easy to generate for masses of order $H$.
For example, the supergravity $\eta$-problem for the inflaton can
be translated to the isocurvature sector to generically expect $m^{2}/H^{2}(t_{k_{0}})\sim O(1)$
during inflation. On the other hand, this does not mean that such
spectra are measurable since any observable must be multiplied by
the background energy density of $\chi$, which tends to dilute away
when $m/H$ is not small (as noted by \cite{Lemoine:2009is}). Still,
there are examples \cite{Kasuya:2009up} of supergravity models generating
measurable, strongly blue spectrum. This will be discussed in depth
in Sec.~\ref{sub:Improvement-of-0904.3800}. In many applications,
$\sqrt{\epsilon}\ll1$ which means that Eq.~(\ref{eq:errorestimate})
imposes a constraint on $|\chi_{0}(t_{k})|$ for the longest wavelengths
leaving the horizon.
\begin{description}
\item [{proof}]~
\end{description}
To make an inflationary prediction based on theorem 2, we need to
match the classical gauge invariant quantity
\begin{equation}
\Delta_{s_{\chi}}^{2}(k)=\frac{k^{3}}{2\pi^{2}}\int\frac{d^{3}k'}{(2\pi)^{3}}\frac{4}{\chi_{0}^{2}}\langle\delta\chi_{nad}(t,\vec{k})\delta\chi_{nad}(t,\vec{k}')\rangle\label{eq:isocurvaturepowerspedc}
\end{equation}
to a quantum correlator computation. A technical difficulty lies in
the fact that $\delta\chi_{nad}(t,\vec{k})$ is not the field that
is being matched to the quantized field (recall it is only a particular
part of a classical solution whose initial condition statistics are
captured by the correlator of Eq.~(\ref{eq:isocurvaturepowerspedc})).
One way to quantize is to quantize the modes of the linearized equation
of motion for $\delta\chi^{(G)}(t,\vec{k})$ in a particular gauge
$G$. It is important to note that we only need the correlator value
computed from field quantization during the few efolds of the horizon
exit since by that time the correlator evolves as a classical statistical
object and $\Delta_{s_{\chi}}^{2}(k)$ is frozen according to theorem
1. If this were not true, we need to understand the full time evolution
of the quantum correlator which is a messy task. Below, we will quantize
in the spatially flat gauge ($G=sf$) instead of the Newtonian gauge
($G=N$) because the gravitational potential does not appear in the
coupled oscillator equations. After quantizing, one can compute the
correlator of Eq.~(\ref{eq:isocurvaturepowerspedc}) by computing
$\langle\delta\chi_{\vec{k}}^{(sf)}\delta\chi_{\vec{p}}^{(sf)}\rangle$
and $\langle\delta\chi_{\vec{k}}^{(sf)}\zeta_{\vec{p}}\rangle$.

Let us now consider the details obtaining Eq.~(\ref{eq:isocurvaturepowerspedc})
by matching to the quantum computation during a few efolds of horizon
exit. Since the relationship between the spatially flat gauge isocurvature
field and the Newtonian gauge field is
\begin{equation}
\delta\chi^{(sf)}(t,\vec{x})=\delta\chi^{(N)}(t,\vec{x})-\frac{\Phi^{(N)}}{H}\dot{\chi}_{0}(t),
\end{equation}
we have
\begin{eqnarray}
\left\langle \frac{\delta\chi_{\vec{k}}^{(sf)}}{\chi_{0}}\frac{\delta\chi_{\vec{p}}^{(sf)}}{\chi_{0}}\right\rangle  & = & \left\langle \frac{\delta\chi_{\vec{k}}^{(N)}}{\chi_{0}}\frac{\delta\chi_{\vec{p}}^{(N)}}{\chi_{0}}\right\rangle +\frac{1}{\chi_{0}^{2}}\left(\frac{\dot{\chi}_{0}}{H}\right)^{2}\left\langle \Phi_{\vec{k}}^{(N)}\Phi_{\vec{p}}^{(N)}\right\rangle \nonumber \\
 &  & -\left(\frac{\dot{\chi}_{0}}{H}\right)\frac{1}{\chi_{0}}2\Re\left\langle \frac{\delta\chi_{\vec{k}}^{(N)}}{\chi_{0}}\Phi_{\vec{p}}^{(N)}\right\rangle .\label{eq:startingeq}
\end{eqnarray}
Because of the linear spectator property, $\Phi_{\vec{p}}^{(N)}$
during inflation is given by the adiabatic mode. We parameterize $\delta\chi_{\vec{p}}^{(N)}$
using the nonadiabatic mode $\delta\chi_{nad}^{(N)}(t,\vec{p})$:
\begin{eqnarray}
\delta\chi_{\vec{p}}^{(N)} & \approx & \delta\chi_{ad}^{(N)}(t,\vec{p})+\delta\chi_{nad}^{(N)}(t,\vec{p})=-\zeta_{\vec{p}}\frac{\dot{\chi}}{a}\int dta+\delta\chi_{nad}^{(N)}(t,\vec{p})\\
\Phi_{\vec{p}}^{(N)} & \approx & \zeta_{\vec{p}}\left[1-\frac{H}{a}\int dta\right].
\end{eqnarray}
We find
\begin{eqnarray}
\left\langle \frac{\delta\chi_{\vec{k}}^{(sf)}}{\chi_{0}}\frac{\delta\chi_{\vec{p}}^{(sf)}}{\chi_{0}}\right\rangle  & \approx & \left\langle \frac{\delta\chi_{nad}^{(N)}(t,\vec{k})}{\chi_{0}}\frac{\delta\chi_{nad}^{(N)}(t,\vec{p})}{\chi_{0}}\right\rangle +\left\langle \zeta_{\vec{k}}\zeta_{\vec{p}}\right\rangle \left(\frac{\dot{\chi}_{0}}{H\chi_{0}}\right)^{2}\nonumber \\
 &  & -2\Re\left\langle \frac{\delta\chi_{nad}^{(N)}(t,\vec{k})}{\chi_{0}}\zeta_{\vec{p}}\right\rangle \left(\frac{\dot{\chi}_{0}}{H\chi_{0}}\right).\label{eq:anotherintermedeq}
\end{eqnarray}

To extract $\left\langle \delta\chi_{nad}^{(N)}(t,\vec{k})\delta\chi_{nad}^{(N)}(t,\vec{p})\right\rangle $
using the spatially flat gauge correlators, we also need an expression
for $\left\langle \delta\chi_{nad}^{(N)}(t,\vec{k})\zeta_{\vec{p}}\right\rangle $
in terms of $\left\langle \delta\chi_{\vec{k}}^{(sf)}\zeta_{\vec{p}}\right\rangle .$
Following a similar procedure as before, this is given by
\begin{equation}
\left\langle \frac{\delta\chi_{\vec{k}}^{(sf)}}{\chi_{0}}\zeta_{\vec{p}}\right\rangle =\frac{1}{\chi_{0}}\left\langle \delta\chi_{nad}^{(N)}(t,\vec{k})\zeta_{\vec{p}}\right\rangle -\left\langle \zeta_{\vec{k}}\zeta_{\vec{p}}\right\rangle \frac{\dot{\chi}_{0}}{H\chi_{0}}.
\end{equation}
Hence, we have arrived at the desired expression
\begin{eqnarray}
4\left\langle \frac{\delta\chi_{nad}^{(N)}(t,\vec{k})}{\chi_{0}}\frac{\delta\chi_{nad}^{(N)}(t,\vec{p})}{\chi_{0}}\right\rangle  & \approx & 4\left\langle \left(\frac{\delta\chi_{\vec{k}}^{(sf)}}{\chi_{0}}+\zeta_{\vec{k}}\left(\frac{\dot{\chi}_{0}}{H\chi_{0}}\right)\right)\right.\nonumber \\
 &  & \left.\left(\frac{\delta\chi_{\vec{p}}^{(sf)}}{\chi_{0}}+\zeta_{\vec{p}}\left(\frac{\dot{\chi}_{0}}{H\chi_{0}}\right)\right)\right\rangle \label{eq:nonadlhs}
\end{eqnarray}
where the left hand side is interpreted as a correlator of stochastic
classical fluctuations and the right hand side is computed at the
quantized level. Note that this expression indicates what is conserved
is effectively the naively expected quantity $\delta\chi^{(sf)}/\chi_{0}$
with the gravitational potential effect subtracted. One may naively
think that the second term of Eq.~(\ref{eq:nonadlhs}) contributes
significantly when $m/H\sim O(1)$. However, we will show below how
this term is exactly canceled out. 

Let us now compute the correlators explicitly in the leading slow-roll
approximation. We recall that in the spatially flat gauge, the mode
equations are homogeneous with off-diagonal mixing
\begin{equation}
\left(\partial_{t}^{2}+3H\partial_{t}+\mathbf{M}^{2}\right)\left(\begin{array}{c}
\delta\varphi^{(sf)}\\
\delta\chi^{(sf)}
\end{array}\right)=0\label{eq:matrixeom}
\end{equation}
which allows one to use adiabatic approximation to quantize the system
since the mass matrix $\mathbf{M}^{2}$ varies sufficiently slowly
in time during slow-roll period for approximate diagonalization. Making
a time\textbf{ }independent rotation to the approximately diagonal
basis, we find the approximately diagonal modes $h_{\varphi,\chi}^{(sf)}$
satisfy 
\begin{equation}
\left(\partial_{t}^{2}+3H\partial_{t}+\mathbf{U}^{\dagger}\mathbf{M}^{2}\mathbf{U}\right)\left(\begin{array}{c}
\delta\varphi_{d}^{(sf)}\\
\delta\chi_{d}^{(sf)}
\end{array}\right)=0\label{eq:diagonalmodeeq}
\end{equation}
where
\begin{eqnarray}
M_{11}^{2} & = & V_{\varphi}''+\frac{3}{M_{p}^{2}}\dot{\varphi}_{0}^{2}-\frac{1}{2}\frac{\dot{\varphi}_{0}^{4}}{M_{p}^{4}H^{2}}+2\frac{V_{\varphi}'\dot{\varphi}_{0}}{M_{p}^{2}H}-\frac{\dot{\varphi}_{0}^{2}\dot{\chi}_{0}^{2}}{2M_{p}^{4}H^{2}}\\
 & \approx & (3\eta_{V}-6\epsilon)H^{2}
\end{eqnarray}
\begin{equation}
M_{12}^{2}=M_{21}^{2}=-\frac{1}{2}\left[-6\frac{\dot{\varphi}_{0}\dot{\chi}_{0}}{M_{p}^{2}}+\frac{\dot{\varphi}_{0}^{3}\dot{\chi}_{0}}{M_{p}^{4}H^{2}}+\frac{\dot{\varphi}_{0}\dot{\chi}_{0}^{3}}{M_{p}^{4}H^{2}}-2\frac{\dot{\varphi}_{0}V_{\chi}'}{M_{p}^{2}H}-2\frac{\dot{\chi}_{0}V_{\varphi}'}{M_{p}^{2}H}\right]\label{eq:mixing}
\end{equation}
\begin{equation}
M_{22}^{2}=V_{\chi}''+\frac{3}{M_{p}^{2}}\dot{\chi}_{0}^{2}-\frac{1}{2}\frac{\dot{\chi}_{0}^{4}}{M_{p}^{4}H^{2}}+2\frac{V_{\chi}'\dot{\chi}_{0}}{M_{p}^{2}H}-\frac{\dot{\varphi}_{0}^{2}\dot{\chi}_{0}^{2}}{2M_{p}^{4}H^{2}}
\end{equation}
and $\mathbf{U}^{\dagger}\mathbf{M}^{2}\mathbf{U}$ is a diagonal
matrix. Let's define
\begin{equation}
\kappa\equiv-\frac{M_{12}^{2}}{M_{22}^{2}}|_{t=0}\approx\frac{-\frac{\dot{\varphi}_{0}V_{\chi}'(\chi)}{M_{p}^{2}H}}{m^{2}[1+O\left(\frac{\chi_{0}^{2}}{M_{P}^{2}}\right)]}=\frac{\sqrt{2\epsilon}\chi_{0}}{M_{p}}\mbox{sgn}(\dot{\varphi}_{0})\label{eq:kappa}
\end{equation}
\begin{equation}
L\equiv-\frac{M_{22}^{2}}{M_{11}^{2}}|_{t=0}\approx-\frac{m^{2}[1+O\left(\frac{\chi_{0}^{2}}{M_{P}^{2}}\right)]}{(3\eta_{V}-6\epsilon)H^{2}}\label{eq:Ldef}
\end{equation}
where $\kappa\ll1$ because of Eq.~(\ref{eq:expansioninfield}).
As far as $L$ is concerned, consider two cases. One case will be
$|L|\gg1$ and the other case will be $|L|\ll1$. The former is the
more interesting case, since the other case is the one which people
have phenomenologically considered most commonly and has been well
established.

\paragraph{Case with $|L|\gg1$}

\begin{equation}
\mathbf{U}_{11}=1-\frac{\kappa^{2}\lambda_{M}^{2}}{2}+\frac{\kappa^{2}\lambda_{M}^{3}}{L}+\lambda_{M}^{4}\left(\frac{11\kappa^{4}}{8}-\frac{3\kappa^{2}}{2L^{2}}\right)+...
\end{equation}
\begin{equation}
\mathbf{U}_{22}=1-\frac{\kappa^{2}\lambda_{M}^{2}}{2}+\frac{\kappa^{2}\lambda_{M}^{3}}{L}+\lambda_{M}^{4}\left(\frac{11\kappa^{4}}{8}-\frac{3\kappa^{2}}{2L^{2}}\right)+...
\end{equation}
\begin{equation}
\mathbf{U}_{12}=-\kappa\lambda_{M}+\frac{\kappa\lambda_{M}^{2}}{L}+\lambda_{M}^{3}\left(\frac{3\kappa^{3}}{2}-\frac{\kappa}{L^{2}}\right)+\lambda_{M}^{4}\left(\frac{\kappa}{L^{3}}-\frac{9\kappa^{3}}{2L}\right)+...
\end{equation}
\begin{equation}
\mathbf{U}_{21}=\kappa\lambda_{M}-\frac{\kappa\lambda_{M}^{2}}{L}-\lambda_{M}^{3}\left(\frac{3\kappa^{3}}{2}-\frac{\kappa}{L^{2}}\right)-\lambda_{M}^{4}\left(\frac{\kappa}{L^{3}}-\frac{9\kappa^{3}}{2L}\right)+...
\end{equation}
with formal perturbation power $\lambda_{M}$ assignment as $M_{11}^{2}(0)=O(\lambda_{M}),\,\, M_{12}^{2}(0)=O(\lambda_{M}),\,\, M_{22}^{2}(0)=O(1)$
and $t=0$ is defined to be initial time here when the modes are deep
within the horizon (i.e. when $k/a(0)\gg H$ ). Note that this diagonalization
differs in a numerically insignificant manner from that used in \cite{Byrnes:2006fr}
where the diagonalization is carried out the horizon exit time (See
Eq.~(8) in \cite{Byrnes:2006fr}).%
\footnote{Since the vacuum physically corresponds to zero on-shell particle
states when the modes are approximately Minkowskian, diagonalization
of the mass matrix when the modes are deep within the horizon is physically
more faithful. On the other hand, as long as the mass matrix is sufficiently
slow in its time dependence, the distinction is not numerically important.
One might also argue that a time dependent diagonalization in the
spirit of WKB approximation is even more appropriate when the modes
are deep within the horizon because as shown in \cite{Bartolo:2001vw},
the time dependent rotation always generates mixing at any time $\dot{\chi}_{0}/\dot{\varphi}_{0}\neq0$.
However, as noticed in \cite{Byrnes:2006fr}, the ratio of the off-diagonal
mass squared and the diagonal mass squared is proportional to slow-roll
parameters whose time variation is suppressed by higher order in slow-roll.
In that sense, neglecting the \emph{time dependence} of the mixing
matrix is justified.%
} Hence, because
\begin{equation}
\frac{\chi_{0}}{M_{p}}\ll1,
\end{equation}
the off-diagonal element of the mixing matrix $\mathbf{U}$ is certainly
generically negligible at the initial time.

Based on the smallness of mixing in $\mathbf{U}$, one might naively
think that the mixing continues to be negligible for the far superhorizon
evolution of $\delta\chi^{(sf)}$ if $\chi_{0}/M_{p}\ll1$ since 
\begin{equation}
\frac{d^{2}\delta\chi^{(sf)}}{dt^{2}}+3H\delta\chi^{(sf)}+\left(\frac{k^{2}}{a^{2}}+M_{22}^{2}\right)\delta\chi^{(sf)}=O(\chi_{0}/M_{p})\delta\varphi^{(sf)}\label{eq:spatiallyflatgaugeeq}
\end{equation}
which may naively allow us to neglect the $M_{21}^{2}$ dependent
right hand side in the limit $\chi_{0}/M_{p}\rightarrow0$. However,
this is not quite right. Since $\kappa\ll1$ and $L\gg1$ which implies
\begin{equation}
\frac{3}{2}H^{2}>M_{22}^{2}\gg M_{11}^{2},M_{12}^{2},\label{eq:masshierarchy}
\end{equation}
the field $\delta\varphi^{(sf)}$ will \emph{eventually} grow relative
to $\delta\chi^{(sf)}$: to see this, use the trial superhorizon positive
frequency mode solution $e^{-i\omega t}$ and consider the resulting
dispersion relationships in the constant $H$ limit. In contrast,
in the decoupling limit, $\delta\varphi^{(sf)}$ will evolve independently
of $\delta\chi^{(sf)}$.%
\footnote{Neglecting $\delta\chi^{(sf)}$ influence on $\delta\varphi^{(sf)}$
is certainly valid since the mixing is small and $\delta\varphi^{(sf)}$
grows relative to $\delta\chi^{(sf)}$ when Eq.~(\ref{eq:masshierarchy})
is satisfied as one can see by considering the positive frequency
mode equations.%
}Because of these two features, the mode $\delta\chi^{(sf)}$ obtains
an effectively inhomogeneous (i.e. sourced) contribution proportional
to $\delta\varphi^{(sf)}$:
\begin{equation}
\delta\chi_{\mbox{particular}}^{(sf)}(t)\approx\frac{\dot{\chi}_{0}(t)}{\dot{\varphi}_{0}(t)}\delta\varphi^{(sf)}(t)\label{eq:secularpiece}
\end{equation}
where we have used slow-roll equations of motion but have\emph{ not}
used the superhorizon approximation.%
\footnote{In spatially flat gauge, this particular solution is related to the
gauge invariant perturbations through $\zeta\approx-H\delta\varphi^{(sf)}/\dot{\varphi}_{0}$.%
} The fact that this is a valid solution in the subhorizon region is
explicitly demonstrated in Appendix \ref{sec:Particular-Solution-In}.
On top of this, one can add a homogeneous solution 
\begin{equation}
\delta\chi^{(sf)}(t)\approx Ch_{\chi}^{(sf)}(t)+\frac{\dot{\chi}_{0}(t)}{\dot{\varphi}_{0}(t)}\delta\varphi^{(sf)}(t)\label{eq:keyeq}
\end{equation}
where $h_{\chi}^{(sf)}$ is a free oscillator field satisfying

\begin{equation}
\frac{d^{2}h_{\chi}^{(sf)}}{dt^{2}}+3Hh_{\chi}^{(sf)}+\left(\frac{k^{2}}{a^{2}}+M_{22}^{2}\right)h_{\chi}^{(sf)}=0
\end{equation}
with the normalization
\begin{equation}
\langle h_{\chi}^{(sf)}(t,\vec{k})h_{\chi}^{(sf)}(t,\vec{p})\rangle\underset{k/a\rightarrow\infty}{\sim}\frac{1}{2ka^{2}}(2\pi)^{3}\delta^{(3)}(\vec{k}+\vec{p})\label{eq:norm_hx}
\end{equation}
and $C$ is a coefficient determined by Bunch-Davies boundary conditions
(note the basis $h$ defined here already satisfies the Bunch-Davies
boundary conditions). Note that if $H$ and $M_{22}$ are constants,
$h_{\chi}^{(sf)}$ is composed of Hankel functions. 

Another way to justify the decomposition into the particular and homogeneous
solution here is that we have to \emph{keep} the second term of Eq.~(\ref{eq:keyeq})
which goes as
\begin{equation}
\frac{\dot{\chi}_{0}(t)}{\dot{\varphi}_{0}(t)}\delta\varphi^{(sf)}(t)\approx\frac{-m^{2}}{3\sqrt{2\epsilon}H^{2}\mbox{sgn}\dot{\varphi}_{0}}\frac{\chi_{0}}{M_{p}}\delta\varphi^{(sf)}(t)\label{eq:enhancedterm}
\end{equation}
even in the decoupling limit of $\chi_{0}/M_{p}\ll1$ because of the
unsuppression due to the relative growing nature of $\delta\varphi^{(sf)}$
relative to $\delta\chi^{(sf)}$. This contribution is small on subhorizon
scales but grows after the modes leave the horizon. This is a type
of secular effect in which long time behavior unsuppresses the naively
suppressed contribution. 

Let's now estimate $C$ in Eq.~(\ref{eq:keyeq}). By definition
\begin{equation}
\delta\chi^{(sf)}=\mathbf{U}_{21}\delta\varphi_{d}^{(sf)}+\mathbf{U}_{22}\delta\chi_{d}^{(sf)}\label{eq:mixed}
\end{equation}
where $\delta\varphi_{d}^{(sf)}$ and $\delta\chi_{d}^{(sf)}$ are
decoupled at the initial time $t=0$, with mass matrix eigenvalues
$m_{1}^{2}$ and $m_{2}^{2}$ respectively (see Eq.~(\ref{eq:diagonalmodeeq})).
To determine $C$, we want to match Eqs.~(\ref{eq:keyeq}) and (\ref{eq:mixed}).
Note that the decoupling used in Eq.~(\ref{eq:keyeq}) already drops
$O(\chi_{0}/M_{p})$ contributions \textbf{except} for those that
can eventually be unsuppressed by the relative growth of $\delta\varphi^{(sf)}$
compared to $\delta\chi^{(sf)}$ after the modes leave the horizon.%
\footnote{Although $\mathbf{U}_{21}\delta\varphi_{d}^{(sf)}$ in Eq.~(\ref{eq:mixed})
is also unsuppressed by $\delta\varphi_{d}^{(sf)}$ in the superhorizon
region, the coefficient $\mathbf{U}_{21}=O(\sqrt{\epsilon}\kappa)$
has an extra power of slow roll parameter $\epsilon$ when compared
to Eq.~(\ref{eq:enhancedterm}). We do not really need this fact
for the demonstration here since we are matching in the subhorizon
region. It is being mentioned to note that the second term of Eq.~(\ref{eq:keyeq})
is not coming from $\mathbf{U}_{21}$ which is something that is being
evaluated only at the initial time.%
} In the subhorizon region, there is no enhancement due to relative
growth of $\delta\varphi^{(sf)}$ to $\delta\chi^{(sf)}$. Since we
will be matching in the subhorizon region, we can drop the $\mathbf{U}_{12}$
and $\mathbf{U}_{21}$ terms when matching: 
\begin{eqnarray}
\delta\chi^{(sf)} & = & Ch_{\chi}^{(sf)}(t)+\frac{\dot{\chi}_{0}(t)}{\dot{\varphi}_{0}(t)}\delta\varphi^{(sf)}(t)\approx\delta\chi_{d}^{(sf)}(t)\label{eq:subhorizonmatch}\\
\delta\phi^{(sf)} & \approx & \delta\phi_{d}^{(sf)}(t)
\end{eqnarray}
At earlier times when the modes are subhorizon, from the normalization
(\ref{eq:norm_hx}) we have 
\begin{equation}
\left\langle \delta\chi_{d}^{(sf)}(t,\vec{k})\delta\chi_{d}^{(sf)}(t,\vec{p})\right\rangle \underset{k/a\rightarrow\infty}{\sim}\left\langle h_{\chi}^{(sf)}(t,\vec{k})h_{\chi}^{(sf)}(t,\vec{p})\right\rangle \label{eq:approxdiagmode}
\end{equation}
if
\begin{equation}
m_{2}^{2}\approx M_{22}^{2}
\end{equation}
consistently with Eq.~(\ref{eq:masshierarchy}). Then Eq.~(\ref{eq:subhorizonmatch})
gives
\begin{eqnarray}
\left|C\right|^{2}\left\langle h_{\chi}^{(sf)}(t,\vec{k})h{}_{\chi}^{(sf)}(t,\vec{p})\right\rangle  & \approx & \left\langle \delta\chi_{d}^{(sf)}(t,\vec{k})\delta\chi_{d}^{(sf)}(t,\vec{p})\right\rangle \nonumber \\
 &  & +\left(\frac{\dot{\chi}_{0}(t)}{\dot{\varphi}_{0}(t)}\right)^{2}\left\langle \delta\mbox{\ensuremath{\phi}}_{d}^{(sf)}(t,\vec{k})\delta\mbox{\ensuremath{\phi}}_{d}^{(sf)}(t,\vec{p})\right\rangle .
\end{eqnarray}
Assuming $\left\langle \delta\chi_{d}^{(sf)}(t,\vec{k})\delta\chi_{d}^{(sf)}(t,\vec{p})\right\rangle \sim\left\langle \delta\phi_{d}^{(sf)}(t,\vec{k})\delta\phi_{d}^{(sf)}(t,\vec{p})\right\rangle $
in the $k/a\to\infty$ limit, we can conclude that 
\begin{equation}
C=1+O\left(\frac{m^{2}}{3\sqrt{2\epsilon}H^{2}}\frac{\chi_{0}}{M_{p}}\right).
\end{equation}

Eq.~(\ref{eq:subhorizonmatch}) thus can be written as 
\begin{equation}
\frac{\delta\chi_{\vec{k}}^{(sf)}}{\chi_{0}}+\zeta_{\vec{k}}\left(\frac{\dot{\chi}_{0}}{H\chi_{0}}\right)=\frac{h_{\chi}^{(sf)}(t,\vec{k})}{\chi_{0}}
\end{equation}
where we used that in spatially flat gauge
\begin{equation}
\zeta=-H\delta\varphi^{(sf)}/\dot{\varphi}_{0}.
\end{equation}
Hence, approximating the Bunch-Davies state correlator near horizon
exit time as 
\begin{equation}
\langle h_{\chi}^{(sf)}(t,\vec{k})h_{\chi}^{(sf)}(t,\vec{p})\rangle=\frac{(2\pi)^{3}}{a^{3}}\frac{\pi}{4}\frac{1}{H}\left|H_{\nu}^{(1)}(\frac{k}{aH})\right|^{2}\delta^{(3)}(\vec{k}+\vec{p})
\end{equation}
where 
\begin{equation}
\nu=\frac{3}{2}\sqrt{1-\frac{4}{9}\frac{m^{2}}{H^{2}}},
\end{equation}
we expand this in the limit $k/(aH)\rightarrow0$ in the usual manner
to arrive at 
\begin{equation}
\left\langle \frac{\delta\chi_{nad}^{(N)}(t,\vec{k})}{\chi_{0}}\frac{\delta\chi_{nad}^{(N)}(t,\vec{p})}{\chi_{0}}\right\rangle \approx(2\pi)^{3}\delta^{(3)}(\vec{k}+\vec{p})\left\{ \frac{1}{a^{3}\chi_{0}^{2}}2^{2(\nu-1)}|\Gamma(\nu)|^{2}\left(\frac{k}{aH}\right)^{-2\nu}\frac{1}{H\pi}\right\} \label{eq:almostfinal}
\end{equation}
which is only valid when 
\begin{equation}
\left|\nu\ln\left(\frac{k}{aH}\right)\right|\gg|\ln f|\label{eq:validity}
\end{equation}
where $f$ is the fractional accuracy desired and $H$ is approximately
constant. Even Eq.~(\ref{eq:validity}) is about justifying keeping
the leading term in the Hankel function expansion, we will soon discuss
that this also corresponds to the attractor behavior needed for the
validity of theorem 1 that will be used here.

Eq.~(\ref{eq:almostfinal}) is not manifestly frozen since for example
the first term seems to dilute as $a^{3}$. However, just as we stated
in theorem 1 (which is more general for classical solutions than this
computation of a few efolds during inflation), this expression does
freeze. To see this, during the few efolds period of horizon exit
during inflation, we can solve
\begin{equation}
\ddot{\chi}_{0}+3H\dot{\chi}_{0}+m^{2}\chi_{0}=0
\end{equation}
to obtain the attractor solution
\begin{equation}
\chi_{0}(t)\approx\chi_{0}(t_{k})\left(\frac{a(t)}{a(t_{k})}\right)^{-\frac{3}{2}+\nu+O(\epsilon)g_{0}(m/H)}\label{eq:chievolution}
\end{equation}
where $k/a(t_{k})=H(t_{k})$, $g_{0}(x)$ is a function which vanishes
in the limit $m/H\rightarrow0$, and the approximation drops the decaying
solution. Hence, one sees that all the $a(t)$ dependencies cancel
in Eq.~(\ref{eq:almostfinal}) and are replaced by $a(t_{k})$.

Let us now consider matching the quantum computation to the classical
variable of theorem 1. With the usual Bunch-Davies normalizations,
$\delta\chi_{nad}$ on superhorizon scales is determined by purely
imaginary mode functions: i.e. the time dependence of mode functions
are determined by by $H_{\nu}^{(1)}(k/(aH))$ where
\begin{equation}
H_{\nu}^{(1)}(x)=J_{\nu}^{(1)}(x)+iY_{\nu}^{(1)}(x)
\end{equation}
 where the $Y_{\nu}^{(1)}(x\rightarrow0)\rightarrow\infty.$ Thus
$\delta\chi_{nad}$ commute like a classical random variables once
$Y_{\nu}^{(1)}$ dominates over $J_{\nu}^{(1)}$. The time scale on
which $Y_{\nu}^{(1)}$ dominates over $J_{\nu}^{(1)}$ is
\begin{equation}
\tau=\frac{1}{2\nu H}.
\end{equation}
This is coincidentally the same time scale as the classical attractor
behavior discussed in theorem 1 and approximately the same time scale
after which Eq.~(\ref{eq:validity}) becomes satisfied. Once 
\begin{equation}
\exp[-(t-t_{k})/\tau]\ll f\label{eq:classical}
\end{equation}
condition is satisfied, one can match the correlator of classical
random variable $S_{\chi}$ to the quantum correlator on superhorizon
scales using Eq.~(\ref{eq:nonadlhs}): 
\begin{equation}
\langle S_{\chi}S_{\chi}\rangle=4(2\pi)^{3}\delta^{(3)}(\vec{k}+\vec{p})\left\{ \frac{1}{a^{3}\chi_{0}^{2}}2^{2(\nu-1)}|\Gamma(\nu)|^{2}\left(\frac{k}{aH}\right)^{-2\nu}\frac{1}{H\pi}\right\} .\label{eq:matchclassicalandquantum}
\end{equation}
In other words, to use theorems 1 and 2 to connect late time isocurvature
to the quantum correlator, we must satisfy Eq.~(\ref{eq:validity}).

Finally, it is instructive to convert Eq.~(\ref{eq:matchclassicalandquantum})
into the form of Eq.~(\ref{eq:result}). We start by using Eq.~(\ref{eq:chievolution})
to recast Eq.~(\ref{eq:matchclassicalandquantum}) as 
\begin{equation}
\Delta_{s_{\chi}}^{2}(k)=4\left[\frac{2^{2\nu}H^{2}|\Gamma(\nu)|^{2}}{\chi_{0}^{2}(t_{k})(2\pi)^{3}}\right]\label{eq:spectrumtoleadingslow-roll}
\end{equation}
where one must keep in mind that despite its appearance, we have already
evaluated this at a time when $a(t)\gg a(t_{k})$ and we have already
assumed Eq.~(\ref{eq:validity}) is satisfied (we will find the constraint
imposed by the assumption below). Choose a time $t_{k_{0}}$ corresponding
to a fiducial mode horizon exit time to write
\begin{equation}
H(t_{k})=H(t_{k_{0}})\left(\frac{k}{k_{0}}\right)^{-\epsilon}
\end{equation}
\begin{equation}
\frac{a(t_{k_{0}})}{a(t_{k})}=\left(\frac{k}{k_{0}}\right)^{-\epsilon-1}
\end{equation}
implying
\begin{equation}
\Delta_{s_{\chi}}^{2}(k)=4\left[\frac{2^{2\nu}H^{2}(t_{k_{0}})|\Gamma(\nu)|^{2}}{\chi_{0}^{2}(t_{k_{0}})(2\pi)^{3}}\right]\left(\frac{k}{k_{0}}\right)^{(3-2\nu)-2\epsilon+O(\epsilon)g_{0}(m/H)}.\label{eq:speceq}
\end{equation}
Note also that we are keeping the $\ln(k/k_{0})$ enhanced $\epsilon$
dependence while dropping other non-enhanced $\epsilon$ dependences.
Since we only specify that $g_{0}(m/H)$ vanishes when $m=0$, the
$-2\epsilon$ power is numerically meaningful in the current estimate
only when $m=0$, which is really not about the blue spectra. Nonetheless,
we keep it here to connect this spectra to the massless axion spectra.

Let us now find the parametric region where one expects the attractor
to be reached consistently with Eq.~(\ref{eq:validity}). An obstacle
to satisfying Eq.~(\ref{eq:validity}) in the context of Eq.~(\ref{eq:spectrumtoleadingslow-roll})
is that either inflation ends too early or the assumption that $H$
is constant during the realization of Eq.~(\ref{eq:classical}) is
invalid. The assumption of constant $H$ breaks down on a time scale
$\Delta t$ satisfying
\begin{equation}
|\dot{H}|\Delta t=fH
\end{equation}
where $f$ is the accuracy desired. During slow-roll, we have 
\begin{equation}
\Delta t=\frac{f}{\epsilon H}.\label{eq:interval}
\end{equation}
Hence, Eq.~(\ref{eq:validity}) becomes
\begin{equation}
\nu\tilde{N}_{k}\gg\left|\ln f\right|
\end{equation}
where $\tilde{N}_{k}\in\{f/\epsilon_{k},\, H(t_{e}-t_{k})\}$ where
$t_{k}$ is the time of the horizon exit of mode $k$ and $t_{e}$
is the end of inflation. 

Note also even if $|\dot{H}|\Delta t/H$ is within error tolerance
$f$, it may still be bigger than what is need to keep $\nu$ real.
Expressed in terms of $\nu$ with $H$ evaluated at the fiducial $k_{0}$
horizon crossing, we need
\begin{equation}
1-\frac{4}{9}\frac{m^{2}}{(H(t_{k_{0}})-\epsilon_{k_{0}}H^{2}(t_{k_{0}})(t_{k}-t_{k_{0}}))^{2}}>0
\end{equation}
which translates to
\begin{equation}
2\varepsilon_{k_{0}}\frac{m^{2}}{H^{2}(t_{k_{0}})\nu^{2}}\ln\left(k/k_{0}\right)<1\,\,\,\,\,\,\mbox{for }k>k_{0}.\label{eq:realityofnu}
\end{equation}

Furthermore, the prefactor terms $2^{\nu}\left|\Gamma(\nu)\right|^{2}$
and $\chi_{0}(t_{k})$ were approximated by expanding about $t_{k_{0}}$
assuming a constant $H$. We require the correction to $2^{\nu}\left|\Gamma(\nu)\right|^{2}$
to be within the error tolerance. This means that in considering the
large enhancement situation $2^{\nu}\left|\Gamma(\nu)\right|^{2}\approx1/\nu^{2}$
near $\nu=0$, we impose 
\begin{equation}
2^{\nu_{k}}\left|\Gamma(\nu_{k})\right|^{2}\approx\frac{1}{\nu^{2}}\left(1+2\varepsilon_{k_{0}}\frac{m^{2}}{H^{2}(t_{k_{0}})\nu^{2}}\ln\left(k/k_{0}\right)\right)<\frac{1}{\nu^{2}}(1+f),
\end{equation}
that gives Eq.~(\ref{eq:constr_nu}). Finally, Eq.~(\ref{eq:interval})
restricts $k$ through the horizon crossing considerations to be in
the range
\begin{equation}
k_{0}\exp\left(-\frac{f}{\epsilon_{k_{0}}}\right)<k<k_{0}\exp\left(\frac{f}{\epsilon_{k_{0}}}\right).
\end{equation}

\paragraph{Case with $|L|\ll1$}

In this case, we have
\begin{equation}
L\equiv-\frac{M_{22}^{2}}{M_{11}^{2}}|_{t=0}\approx-\frac{m^{2}[1+O\left(\frac{\chi^{2}}{M_{P}^{2}}\right)]}{(3\eta_{V}-6\epsilon)H^{2}}\ll1.
\end{equation}
Because we still have $\chi_{0}/M_{p}\ll1$ and
\begin{equation}
M_{21}^{2}\approx\frac{m^{2}\chi_{0}\sqrt{2\epsilon}}{M_{p}}\mbox{sgn}\dot{\varphi}_{0},
\end{equation}
we have the hierarchy
\begin{equation}
M_{12}^{2}=M_{21}^{2}\ll M_{22}^{2}\ll M_{11}^{2}.\label{eq:otherhierarchy}
\end{equation}
Hence, the off-diagonal element of the mixing matrix $\mathbf{U}$
is still suppressed. (In fact, it is now suppressed relative to all
of the matrix elements.) Furthermore, again because of the mass hierarchy
in Eq.\ (\ref{eq:otherhierarchy}), $\delta\varphi^{(sf)}$ does
not grow relative to $\delta\chi^{(sf)}$ unlike the situation explained
just after Eq.~(\ref{eq:spatiallyflatgaugeeq}). Hence, we can in
this case completely ignore the mixing.

Finally, we see the term shifting the $\delta\chi^{(sf)}$ field in
Eq.~(\ref{eq:nonadlhs}) is negligible to leading slow-roll order
accuracy: i.e. 
\begin{equation}
\zeta_{\vec{k}}\left(\frac{\dot{\chi}_{0}}{H\chi}\right)\approx\zeta_{\vec{k}}\left(\frac{-m^{2}}{3H^{2}}\right)\ll(3\eta_{V}-6\epsilon)\zeta_{k}.
\end{equation}
Hence, we again conclude that the isocurvature spectrum is given by
Eq.~(\ref{eq:spectrumtoleadingslow-roll}). 

Just as for theorem 1, a trivial corollary of this theorem is to discuss
the situation when the constant $m$ is replaced by $m(t)$ which
is constant during a finite time interval during inflation and makes
a transition to another value during inflation. 
\begin{description}
\item [{corollary~2}] In the context of theorem 3 and just as in corollary
1, suppose $m$ is not a constant but makes a transition to another
value during inflation:
\begin{equation}
m^{2}(t)=\begin{cases}
m_{1} & \,\,\,\,\,\, t<t_{c}\\
m_{2} & \,\,\,\,\,\, t>t_{c}
\end{cases}
\end{equation}
where the transition time region near $t=t_{c}$ is assumed to be
much smaller in time than $m_{1}^{-1}$ and $H^{-1}$. Suppose sufficient
time has passed during the $t<t_{c}$ period to be in the attractor
approximation just as in corollary 1: i.e.
\begin{equation}
\tilde{N}(t_{c},t_{k})\nu(m_{1})>|\ln f|\label{eq:tildeNconclusion}
\end{equation}
where $\tilde{N}(t_{c},t_{k})\in\{f/\epsilon_{k},\,(t_{c}-t_{k})H\},$
\begin{equation}
\nu(m_{1})=\frac{3}{2}\sqrt{1-\frac{4}{9}\frac{m_{1}^{2}}{H^{2}},}\label{eq:timedependentnufunc}
\end{equation}
and $f<1$ is the error tolerance in the computation. The spectrum
is still given by Eq.~(\ref{eq:mainfinresult}) with $\nu\rightarrow\nu(m_{1})$
for modes $k$ in the range
\begin{equation}
k_{\mbox{min}}<k<k_{\mbox{max}}
\end{equation}
where $k_{\mbox{max}}$ is the smallest $k$ that among \{$k$ that
saturates the inequality of Eq.~(\ref{eq:tildeNconclusion}), $k_{0}\exp(f/\epsilon_{k_{0}})$,
$k_{0}\exp\left[\frac{\nu^{2}}{2\epsilon_{k_{0}}}\frac{H^{2}(t_{k_{0}})}{m^{2}}\right]$\}
and 
\begin{equation}
k_{\mbox{min}}=\max\left\{ k_{0}\exp\left(-\frac{f}{\epsilon_{k_{0}}}\right),\, a(t_{b})H(t_{b})\right\} 
\end{equation}
where $t_{b}$ is the beginning of inflation. The fractional error
$O(\mathcal{E})$ in Eq.~(\ref{eq:mainfinresult}) receives contributions
from
\begin{equation}
\mathcal{E}=\exp\left[-2\nu(m_{1})(t_{c}-t_{k})H\right]+\frac{\delta\rho_{\chi}^{(N)}}{\delta\rho_{\mbox{dominant}}^{(N)}}+\frac{m^{2}}{3\sqrt{2\epsilon}H^{2}}\frac{|\chi_{0}(t_{k})|}{M_{p}}+\frac{|\chi_{0}(t_{k})|}{M_{p}}+\epsilon_{k_{0}}+f.\label{eq:errorestimaterepeat}
\end{equation}

\end{description}
The error contribution proportional to $1/\sqrt{\epsilon}$ in Eq.~(\ref{eq:errorestimaterepeat})
can be rewritten as abound on $\chi_{0}(t_{k_{0}})/M_{p}$ if we require
it to be less than $f$ and require $k_{H_{0}}\approx a_{0}H_{0}\pi/2$%
\footnote{$k_{H_{0}}$ is the wave number corresponding to the observable universe
today that in a comoving box with a length of $L=4/a_{0}H_{0}$.%
} mode to be the quantizable within the current analytic treatment:
\begin{equation}
\frac{2}{3}\pi\left(\frac{k_{H_{0}}}{k_{0}}\right)^{-\frac{3}{2}+\nu+O(\epsilon)g_{0}(m/H)}\frac{m^{2}}{H^{2}}\sqrt{\Delta_{\zeta}^{2}(k_{H_{0}})}\frac{M_{p}}{H}\frac{1}{f}<\frac{M_{p}}{|\chi_{0}(t_{k_{0}})|}.\label{eq:nontrivial}
\end{equation}
This and Eq.~(\ref{eq:mixturefirst}) give
\begin{equation}
\frac{\Delta_{s}^{2}(k_{H_{0}})}{\Delta_{\zeta}^{2}(k_{H_{0}})}=\omega_{\chi}^{2}\frac{\Delta_{s_{\chi}}^{2}(k_{H_{0}})}{\Delta_{\zeta}^{2}(k_{H_{0}})}>\omega_{\chi}^{2}\frac{2^{2\nu-1}|\Gamma(\nu)|^{2}}{\pi}\left[\frac{2}{3f}\frac{m^{2}}{H^{2}}\right]^{2}.\label{eq:generalamplitudebound}
\end{equation}
Since we know that the left hand side is constrained to a few percent
level (at least for the scale invariant spectra), phenomenology requires
that the dark matter fraction in $\chi$ be very small for a very
blue spectra: i.e. $\omega_{\chi}\ll1$ for $m/H\sim O(1)$. 
\begin{description}
\item [{proof}]~
\end{description}
For those modes which can have Bunch-Davies boundary conditions at
early time and exits the horizon long before $t=t_{c}$ will lead
to a spectrum where Eqs.~(\ref{eq:almostfinal}) and (\ref{eq:validity})
are valid with $\nu(m)\rightarrow\nu(m_{1})$:
\begin{equation}
\left\langle \frac{\delta\chi_{nad}^{(N)}(t,\vec{k})}{\chi_{0}}\frac{\delta\chi_{nad}^{(N)}(t,\vec{p})}{\chi_{0}}\right\rangle \approx(2\pi)^{3}\delta^{(3)}(\vec{k}+\vec{p})\left\{ \frac{2^{2(\nu(m_{1})-1)}}{a^{3}\chi_{0}^{2}}\frac{|\Gamma(\nu(m_{1}))|^{2}}{H\pi}\left(\frac{k}{aH}\right)^{-2\nu(m_{1})}\right\} \label{eq:alternatealmostfinal}
\end{equation}
\begin{equation}
\nu(x)\equiv\frac{3}{2}\sqrt{1-\frac{4}{9}\frac{x^{2}}{H^{2}}}.
\end{equation}
If the dominance of $Y_{\nu(m_{1})}(x)$ over $J_{\nu(m_{1})}(x)$
occurs as 
\begin{equation}
\exp[-(t-t_{k})2\nu(m_{1})H]\ll f,\label{eq:conditionbeforetransition}
\end{equation}
with $t<t_{c}$, classical behavior is heuristically justified, and
one can match the quantum computation to the classical solution. Since
the constant $H$ approximation result of Eq.~(\ref{eq:alternatealmostfinal})
requires $t<t_{k}+f/(\epsilon_{k}H)$ and Eq.~(\ref{eq:conditionbeforetransition})
must occur with $t<t_{c}$, we arrive at the conclusion of Eq.~(\ref{eq:tildeNconclusion}).
Also, for $\nu$ to be real throughout the time period between when
the mode $k>k_{0}$ leaves the horizon and the mode $k_{0}$ leaves
the horizon, we must have
\begin{equation}
\nu(m_{1})>\sqrt{2\epsilon_{k_{0}}(N_{k}-N_{k_{0}})}\frac{m}{H(k_{0})}\,\,\,\,\,\,\mbox{for }k>k_{0}.\label{eq:alternaterealityofnu}
\end{equation}
which follows from the justification of Eq.~(\ref{eq:realityofnu})
with $\nu\rightarrow\nu(m_{1}).$ This sets an upper bound of $k$
to be at
\begin{equation}
k<k_{0}\exp\left[\frac{\nu^{2}}{2\epsilon_{k_{0}}}\frac{H^{2}(t_{k_{0}})}{m^{2}}\right].
\end{equation}
The $k$ bound coming from $\chi(t_{k})$ being connected to $\chi(t_{k_{0}})$
through a de Sitter space solution is the same as Eq.~(\ref{eq:connectedthroughdS}).

\section{\label{sec:Applications}Applications}

\subsection{\label{sub:Improvement-of-0904.3800}Improvement of the Axion Blue
Isocurvature Scenario \cite{Kasuya:2009up}}

In this section we apply our theorems to the scenario of \cite{Kasuya:2009up}
and compute $O(1/(n-4)^{2})$ corrections to their blue spectrum.

\subsubsection{A Review of the Axion Blue Isocurvature Scenario \cite{Kasuya:2009up}}

We begin by reviewing \cite{Kasuya:2009up}. They consider a renormalizable
superpotential
\begin{equation}
W=h(\Phi_{+}\Phi_{-}-F_{a}^{2})\Phi_{0}
\end{equation}
where the subscripts on $\Phi$ indicate $U(1)$ global charges. The
F-term potential is
\begin{equation}
V_{F}=h^{2}|\Phi_{+}\Phi_{-}-F_{a}^{2}|^{2}+h^{2}(|\Phi_{+}|^{2}+|\Phi_{-}|^{2})|\Phi_{0}|^{2}.
\end{equation}
A flat directions of $V_{F}$ exists along
\begin{equation}
\Phi_{+}\Phi_{-}=F_{a}^{2}\,\,\,\,\,\,\,\,\,\,\,\,\Phi_{0}=0.\label{eq:flatdirection}
\end{equation}
Their soft-SUSY breaking terms are assumed to be
\begin{equation}
V_{soft}=m_{+}^{2}|\Phi_{+}|^{2}+m_{-}^{2}|\Phi_{-}|^{2}+m_{0}^{2}|\Phi_{0}|^{2}
\end{equation}
where $m_{i}=O(\mbox{TeV})$. The Kaehler potential induced potential
is
\begin{equation}
V_{K}=c_{+}H^{2}|\Phi_{+}|^{2}+c_{-}H^{2}|\Phi_{-}|^{2}+c_{0}H^{2}|\Phi_{0}|^{2}
\end{equation}
where $c_{+,-,0}$ are positive $O(1)$ constants. In addition to
these, there can be $H$ induced trilinear terms which can spoil the
flat direction. Hence, they assume that the inflaton sector can be
arranged to have $H\ll F_{a}$ such that the flat directions are only
lifted by the quadratic terms.

Looking along the flat direction of Eq.~(\ref{eq:flatdirection})
(more explicitly, setting $\Phi_{0}=0$), they have the effective
potential being
\begin{equation}
V\approx h^{2}|\Phi_{+}\Phi_{-}-F_{a}^{2}|^{2}+c_{+}H^{2}|\Phi_{+}|^{2}+c_{-}H^{2}|\Phi_{-}|^{2}.
\end{equation}
During inflation, the minimum of $\Phi_{\pm}$ lies at
\begin{equation}
|\Phi_{\pm}^{\mbox{min}}|\approx\left(\frac{c_{\mp}}{c_{\pm}}\right)^{1/4}F_{a}.
\end{equation}
They assume $\Phi_{\pm}$ starts out away from the minimum with a
magnitude larger than this and approaches the minimum during inflation.
This implies the $U(1)$ symmetry is broken during inflation. Hence,
there will be a linear combination of the phases of $\Phi_{\pm}$
which will be the Nambu-Goldstone boson associated with the broken
$U(1)$. Hence, they make a judicious sigma model parameterization
\begin{equation}
\Phi_{\pm}\equiv\frac{\varphi_{\pm}}{\sqrt{2}}\exp\left(i\frac{a_{\pm}}{\sqrt{2}\varphi_{\pm}}\right)\label{eq:angularparam}
\end{equation}
where $\varphi_{\pm}$ and $a_{\pm}$ are real. For our explanation
later, keep in mind that $\varphi_{\pm}$ and $a_{\pm}$ are four
distinct dynamical degrees of freedom.

The potential in the new variable is 
\begin{eqnarray}
V & \approx & -h^{2}F_{a}^{2}\varphi_{+}\varphi_{-}\cos\left[\frac{a_{+}\varphi_{-}+a_{-}\varphi_{+}}{\sqrt{2}\varphi_{+}\varphi_{-}}\right]+h^{2}F_{a}^{4}+\frac{1}{4}h^{2}\varphi_{-}^{2}\varphi_{+}^{2}\nonumber \\
 &  & +\frac{1}{2}c_{+}H^{2}\varphi_{+}^{2}+\frac{1}{2}c_{-}H^{2}\varphi_{-}^{2}\label{eq:potential}
\end{eqnarray}
For any fixed value of $\varphi_{\pm}$, one can see that a linear
combination of $a_{\pm}$ will be the Nambu-Goldstone boson with an
approximately flat direction. That axion combination called $a$ is
determined in their equation 15: 
\begin{equation}
a=\frac{\varphi_{+}}{\sqrt{\varphi_{+}^{2}+\varphi_{-}^{2}}}a_{+}-\frac{\varphi_{-}}{\sqrt{\varphi_{+}^{2}+\varphi_{-}^{2}}}a_{-}
\end{equation}
which is not to be confused with the scale factor (whenever it is
not clear from the context, we will add the subscript $a_{\mbox{scale}}$
to denote the metric scale factor). The orthogonal combination
\begin{equation}
b=\frac{\varphi_{-}}{\sqrt{\varphi_{+}^{2}+\varphi_{-}^{2}}}a_{+}+\frac{\varphi_{+}}{\sqrt{\varphi_{+}^{2}+\varphi_{-}^{2}}}a_{-}\label{eq:biszero}
\end{equation}
has a potential
\begin{equation}
V_{b}=-h^{2}F_{a}^{2}\varphi_{+}\varphi_{-}\cos\left(\frac{\sqrt{\varphi_{+}^{2}+\varphi_{-}^{2}}}{\varphi_{+}\varphi_{-}}b\right).
\end{equation}

They assume $\varphi_{+}$ is initially large (of order of $M_{P}$).
Note that since $\varphi_{+}\varphi_{-}\approx2F_{a}^{2}$ along the
flat direction, if $\varphi_{+}(t_{i})\sim M_{P}$, then
\begin{equation}
\varphi_{-}(t_{i})\sim\frac{F_{a}^{2}}{M_{P}}\ll\varphi_{+}.\label{eq:largevev}
\end{equation}
Hence, the field $b$ during this time near $t_{i}$ has a mass of
order $h\varphi_{+}\sim hM_{P}\gg H$ (for $F_{a}\lesssim M_{P}$)
and thus is assumed to be settled to the minimum of $b=0$ during
inflation. The mass squared matrix of $\varphi_{\pm}$ also says that$\varphi_{+}$
mass squared is of order of $H^{2}$ such that it can be dynamical.
Since $b$ is sitting at $b=0$, Eq.~(\ref{eq:biszero}) implies
\begin{equation}
\delta a_{-}\approx-\frac{\varphi_{-}}{\varphi_{+}}\delta a_{+}.
\end{equation}
Hence, the angles appearing in Eq.~(\ref{eq:angularparam}) are
\begin{equation}
\delta\theta_{+}\equiv\frac{\delta a_{+}}{\sqrt{2}\varphi_{+}}
\end{equation}
 and 
\begin{equation}
\delta\theta_{-}\equiv\frac{\delta a_{-}}{\sqrt{2}\varphi_{-}}
\end{equation}

Because of Eq.~(\ref{eq:largevev}), we have the fluctuations in
the $a$ field being
\begin{equation}
a\approx a_{+}=\theta_{+}\sqrt{2}\varphi_{+}.
\end{equation}
They therefore claim that the quantum fluctuations of 
\begin{equation}
S=\frac{2\delta a}{a}
\end{equation}
is frozen upon horizon departure. They assume $a$ is massless and
assign $\delta a$ an amplitude of $H/(2\pi)$ where $H$ is approximately
constant during inflation. With that reasoning, they write
\begin{equation}
S_{\mbox{amplitude}}=\frac{H}{\sqrt{2}\pi\varphi_{+}\theta_{+}}.\label{eq:theirisocurvature}
\end{equation}
It is important to note here that as far as identifying $\delta a$
with $H/(2\pi)$ is concerned, they are neglecting the kinetic term
induced mass of the $a$ field which is what we are going to correct
below.

Note $\varphi_{+}$decreases as a function of time as
\begin{equation}
\varphi_{+}(t)\approx\varphi_{+}(t_{1})\left(\frac{a_{\mbox{scale}}(t)}{a_{\mbox{scale}}(t_{1})}\right)^{-\frac{3}{2}+\nu}
\end{equation}
 with
\begin{equation}
\nu=\frac{3}{2}\sqrt{1-\frac{4}{9}\frac{m_{++}^{2}}{H^{2}}}
\end{equation}
where $m_{++}^{2}$ is the mass squared of the $\varphi_{+}$ field
and $t_{1}$ being a fiducial time. Since horizon exit condition gives
\begin{equation}
\frac{a_{\mbox{scale}}(t_{1})}{a_{\mbox{scale}}(t)}=\left(\frac{k}{a_{\mbox{scale}}(t_{1})H(t_{1})}\right)^{-\epsilon-1}
\end{equation}
which gives
\begin{equation}
S_{\mbox{amplitude}}=\frac{H(t_{1})}{\sqrt{2}\pi\varphi_{+}(t_{1})\theta_{+}}\left(\frac{k}{a_{\mbox{scale}}(t_{1})H(t_{1})}\right)^{\left(\frac{3}{2}-\nu\right)\left(1+\epsilon\right)-\epsilon}.
\end{equation}
This is indeed a blue spectrum in the limit $\nu\rightarrow0$ (i.e.
for $m/H\rightarrow3/2$).

\subsubsection{Improvement}

According to our theorem 1, what needs to be frozen is $\delta a/a$.
Since $a$ is decaying with time (in contrast to the standard axion
scenario), $\delta a$ cannot be massless. On the other hand, \cite{Kasuya:2009up}
assigns a massless spectral amplitude to $\delta a$ of $H/(2\pi)$
(see Eq.~(\ref{eq:theirisocurvature}) above). The goal of this subsection
is to remedy this. 

Although the action is a bit simpler if we choose other $\sigma$-model
parameterization, we will consider the explicit axion parameterization
$\{a,b,\varphi_{\pm}\}$ considered in \cite{Kasuya:2009up} for transparency
in connecting with this work. The kinetic term term is
\begin{eqnarray}
K & = & |\partial\Phi_{+}|^{2}+|\partial\Phi_{-}|^{2}\label{eq:kineticdef}\\
 & = & \frac{1}{4}(\partial a)^{2}+\frac{1}{4}(\partial b)^{2}+\nonumber \\
 &  & -\frac{a}{2}\partial_{\mu}a\left[{\color{black}{\color{red}{\color{black}\sin^{2}\gamma}{\color{black}\partial^{\mu}\ln\varphi_{-}}}+{\color{red}{\color{black}\cos^{2}\gamma\partial^{\mu}\ln\varphi_{+}}}}\right]+b\partial_{\mu}a\left[{\color{black}{\color{black}{\color{red}{\color{black}\partial^{\mu}\ln\varphi_{-}}}}{\color{red}{\color{black}-\partial^{\mu}\ln\varphi_{+}}}}\right]\sin\gamma\cos\gamma\nonumber \\
 &  & -\frac{b}{2}\partial_{\mu}b\left[{\color{red}{\color{black}\cos^{2}\gamma}{\color{black}\partial^{\mu}\ln\varphi_{-}}}+{\color{red}{\color{black}\sin^{2}\gamma\partial^{\mu}\ln\varphi_{+}}}\right]+\nonumber \\
 &  & \partial_{\mu}\varphi_{-}\partial^{\mu}\varphi_{-}\left[\frac{1}{2}+\frac{a^{2}}{4\varphi_{+}^{2}}\cos^{2}\gamma\sin^{2}\gamma-\frac{ab}{\varphi_{+}^{2}}\cos^{3}\gamma\sin\gamma+\frac{b^{2}}{4\varphi_{+}^{2}}\left(\cos^{2}\gamma+4\sin^{2}\gamma\right)\frac{\cos^{4}\gamma}{\sin^{2}\gamma}\right]+\nonumber \\
 &  & \frac{1}{2}\partial_{\mu}\varphi_{-}\partial^{\mu}\varphi_{+}\frac{1}{\varphi_{+}^{2}}\left[-2ab\cos^{4}\gamma+a^{2}\cos^{3}\gamma\sin\gamma-3b^{2}\cos^{3}\gamma\sin\gamma+2ab\cos^{2}\gamma\sin^{2}\gamma\right]+\nonumber \\
 &  & \partial_{\mu}\varphi_{+}\partial^{\mu}\varphi_{+}\left[\frac{1}{2}+\frac{a^{2}}{4\varphi_{+}^{2}}\cos^{4}\gamma+\frac{ab}{\varphi_{+}^{2}}\cos^{3}\gamma\sin\gamma+\frac{b^{2}}{4\varphi_{+}^{2}}\left(\sin^{2}\gamma+4\cos^{2}\gamma\right)\sin^{2}\gamma\right]\label{eq:kineticful}
\end{eqnarray}
where 
\begin{equation}
\cos\gamma\equiv\frac{\varphi_{+}}{\sqrt{\varphi_{+}^{2}+\varphi_{-}^{2}}}.
\end{equation}
Because $b$ and $\varphi_{-}$ have masses much larger than the expansion
rate $H$, they will sit at the minimum. Although the global minimum
of this potential is at
\begin{equation}
b=0,\,\,\,\,\varphi_{+}|_{min}=\sqrt{2}\sqrt{\frac{\sqrt{c_{-}}}{\sqrt{c_{+}}}F_{a}^{2}-\frac{c_{-}}{h^{2}}H^{2}},\,\,\,\,\,\,\,\varphi_{-}|_{min}=\sqrt{2}\sqrt{\frac{\sqrt{c_{+}}}{\sqrt{c_{-}}}F_{a}^{2}-\frac{c_{+}}{h^{2}}H^{2}}\label{eq:globalmin}
\end{equation}
 assuming $F_{a}$ is sufficiently larger than $H$, when $\varphi_{+}$
is displaced from its global minimum, the $\varphi_{-}$ will sit
at the $\varphi_{-}$ variation minimum of
\begin{equation}
\varphi_{-}=\frac{2F_{a}^{2}}{\varphi_{+}}\frac{1}{1+2\frac{c_{-}}{h^{2}}\frac{H^{2}}{\varphi_{+}^{2}}}
\end{equation}
which varies as a function of time because $\varphi_{+}$ varies as
a function of time. On the other hand, because its mass is heavy,
its fluctuations are not dynamically important. 

For later discussion of the dynamics of the $\varphi_{+}$ and $a$,
it is useful to note that the $a$ field here corresponds to coset
space parameterization of the spontaneously broken PQ symmetry: $U(1)_{PQ}$
defined as
\begin{equation}
\Phi_{\pm}\rightarrow e^{\pm i\theta}\Phi_{\pm}.
\end{equation}
Because $U(1)_{PQ}$ breaking VEV is considered in the dynamics (i.e.
$\varphi_{\pm}$ dynamics), the full dynamics is characterized by
an unbroken $U(1)_{PQ}$. $U(1)_{PQ}$ gives rise to an exactly conserved
current $J_{PQ}^{\mu}$ that is conveniently expressed in terms of
$a_{\pm}$:
\begin{equation}
J_{PQ}^{\mu}=\varphi_{-}\partial^{\mu}a_{-}-\varphi_{+}\partial^{\mu}a_{+}+a_{+}\partial^{\mu}\varphi_{+}-a_{-}\partial^{\mu}\varphi_{-}.
\end{equation}
This current is one piece of crucial information not discussed in
\cite{Kasuya:2009up}.

Integrating out $b$ (i.e. set $b=0$ in Eq.~(\ref{eq:kineticful})),
we have
\begin{eqnarray}
K & \approx & \frac{1}{4}(\partial a)^{2}-\frac{a}{2}\partial_{\mu}a\left[{\color{black}{\color{red}{\color{black}\sin^{2}\gamma}{\color{black}\partial^{\mu}\ln\varphi_{-}}}+{\color{red}{\color{black}\cos^{2}\gamma\partial^{\mu}\ln\varphi_{+}}}}\right]+\nonumber \\
 &  & \partial_{\mu}\varphi_{-}\partial^{\mu}\varphi_{-}\left[\frac{1}{2}+\frac{a^{2}}{4\varphi_{+}^{2}}\cos^{2}\gamma\sin^{2}\gamma\right]+\nonumber \\
 &  & \frac{1}{2}\partial_{\mu}\varphi_{-}\partial^{\mu}\varphi_{+}\frac{1}{\varphi_{+}^{2}}\left[a^{2}\cos^{3}\gamma\sin\gamma\right]+\nonumber \\
 &  & \partial_{\mu}\varphi_{+}\partial^{\mu}\varphi_{+}\left[\frac{1}{2}+\frac{a^{2}}{4\varphi_{+}^{2}}\cos^{4}\gamma\right].\label{eq:firstintermed}
\end{eqnarray}
Next, we will focus on the dynamics while $\varphi_{+}\gg\varphi_{-}$
and $\varphi_{+}\gg a$ where we will be able to integrate out $\varphi_{-}$.
Since $\gamma\ll1$ in this regime, we can then drop the $\sin\gamma$
terms to write
\begin{eqnarray}
K & \approx & \frac{1}{4}(\partial a)^{2}-\frac{a}{2\varphi_{+}}\partial_{\mu}a{\color{black}{\color{red}{\color{black}\partial^{\mu}\varphi_{+}}}}+\nonumber \\
 &  & \frac{1}{2}\partial_{\mu}\varphi_{-}\partial^{\mu}\varphi_{-}+\left[\frac{1}{2}+\frac{a^{2}}{4\varphi_{+}^{2}}\right]\partial_{\mu}\varphi_{+}\partial^{\mu}\varphi_{+}.\label{eq:simplified}
\end{eqnarray}
Next, since the mass of $\varphi_{-}$ is of the order
\begin{equation}
h^{2}\varphi_{+}^{2}\gg H^{2},
\end{equation}
we can integrate out $\varphi_{-}$, leaving $a$ and $\varphi_{+}$
as only dynamical degrees of freedom. Note also that from the consideration
of the kinetic term Eq.~(\ref{eq:simplified}), the axion is no longer
shift invariant as long as $\partial^{\mu}\varphi_{+}\neq0.$ Note
also that the simplification embodied in Eq.~(\ref{eq:simplified})
breaks down when $\varphi_{+}$ is near its global minimum of Eq.~(\ref{eq:globalmin}).
Also, in order to stabilize $\varphi_{+}$ at the minimum, we have
implicitly assumed
\begin{equation}
F_{a}^{2}>\frac{\sqrt{c_{-}c_{+}}}{h^{2}}H^{2}\label{eq:stabilize}
\end{equation}
and in order to decouple $\varphi_{-}$ and $b$ at the minimum, we
have assumed
\begin{equation}
F_{a}^{2}\gg\frac{\sqrt{c_{-}c_{+}}}{h^{2}(c_{-}+c_{+})}H^{2}.\label{eq:keepatmin}
\end{equation}

The equation of motion for $a$ and $\varphi_{+}$ are
\begin{equation}
\square a-a\frac{\square\varphi_{+}}{\varphi_{+}}=0\label{eq:axionmassistimedependent}
\end{equation}
\begin{equation}
\square\varphi_{+}+\frac{1}{\sqrt{g}}\partial_{\mu}\left[\frac{a^{2}}{2\varphi_{+}^{2}}\sqrt{g}\partial^{\mu}\varphi_{+}\right]-\frac{1}{2\varphi_{+}}\frac{1}{\sqrt{g}}\partial_{\mu}\left[\sqrt{g}a\partial^{\mu}a\right]+\partial_{\varphi_{+}}V=0
\end{equation}
Next, consider the background and linearized equations for $a$ and
$\varphi_{+}$ with the $\varphi_{-}$ and $b$ sitting at its minimum
\begin{equation}
b=0
\end{equation}
\begin{equation}
\varphi_{-}=\frac{2F_{a}^{2}}{\varphi_{+}}\left[\frac{1}{1+2\frac{c_{-}}{h^{2}}\frac{H^{2}}{\varphi_{+}^{2}}}\right]
\end{equation}
leading to the approximate potential of 
\begin{equation}
V|_{\varphi_{-},b=min}=\frac{H^{2}}{2}\left(c_{+}\varphi_{+}^{2}+\frac{4c_{-}h^{2}F_{a}^{4}}{h^{2}\varphi_{+}^{2}+2c_{-}H^{2}}\right).\label{eq:Vphiplus}
\end{equation}

With this truncation, we have the conserved $U(1)_{PQ}$ charge density
being
\begin{equation}
J_{PQ}^{0}\approx-(\frac{4F_{a}^{4}}{\varphi_{+}^{4}}+1)\left(\varphi_{+}\dot{a}_{+}-a_{+}\dot{\varphi}_{+}\right).
\end{equation}
Because of the spacetime expansion, this charge density dilutes away:
i.e.
\begin{equation}
\frac{a_{+}(t)}{a_{+}(t_{1})}\approx\frac{\varphi_{+}(t)}{\varphi_{+}(t_{1})}\label{eq:constant}
\end{equation}
to exponential accuracy. Note that during the early period of inflation
when $\varphi_{+}\gg F_{a}$, we have $a\approx a_{+}$. Eq.~(\ref{eq:constant})
and theorem 1 explains why during this time period
\begin{equation}
\frac{\delta a}{a(t)}\approx\frac{\varphi_{+}(t_{1})}{a_{+}(t_{1})}\frac{\delta a}{\varphi_{+}(t)}=\mbox{constant.}
\end{equation}
In other words, $\delta a/\varphi_{+}$ is frozen because it behaves
like $\delta a/a$ because of $U(1)_{PQ}$ even though the numerator
and the denominator are independent degrees of freedom for which theorem
1 would not always apply. For the complete spectrum calculation, it
is better to rewrite Eq.~(\ref{eq:constant}) in terms of $a(t)$:
\begin{equation}
\frac{a(t)}{a(t_{1})}=\frac{\sqrt{\frac{4F_{a}^{4}}{\varphi_{+}^{2}(t)}+\varphi_{+}^{2}(t)}}{\sqrt{\frac{4F_{a}^{4}}{\varphi_{+}^{2}(t_{1})}+\varphi_{+}^{2}(t_{1})}}.
\end{equation}

Let's consider the magnitude of the initial charge that becomes diluted
to assess the accuracy of Eq.~(\ref{eq:constant}): 
\begin{equation}
Q\equiv\int d^{3}xa^{3}(t_{0})J_{PQ}^{0}(t_{0}).
\end{equation}
If we assume all dynamical scales are tied to $H$, we can estimate
\begin{equation}
Q\approx O(H)\varphi_{+}(t_{0})a_{+}(t_{0})a_{\mbox{scale}}^{3}(t_{0})
\end{equation}
and the diluted charge density at any time $t$ after the initial
time $t_{0}$ is
\begin{equation}
J_{PQ}^{0}(t)\sim O(H)\varphi_{+}(t_{0})a_{+}(t_{0})\frac{a^{3}(t_{0})}{a^{3}(t)}.
\end{equation}
This translates into an Eq.~(\ref{eq:constant}) accuracy of 
\begin{equation}
\mbox{accuracy}\equiv O(e^{-3\Delta N})\label{eq:accuracy}
\end{equation}
 where $\Delta N>0$ is the efold time from the beginning of the inflation
$t_{0}$ to the time of horizon exit of a given mode. For example,
if we want an accuracy of $O(\epsilon)$, we only need the beginning
of inflation and the time of horizon exit for the longest wavelength
mode of phenomenological interest labeled by $k_{\mbox{min}}$ to
be separated by the efold number of 
\begin{equation}
\Delta N\sim-\frac{1}{3}\ln\epsilon
\end{equation}
before we can set $t_{1}$ of Eq.~(\ref{eq:constant}) to $t_{k_{\mbox{min}}}$
while assuming for $t>t_{k_{\mbox{min}}}$. For $\epsilon=0.01$,
one only gives up about 1 efolding. During this one efolding, the
$\varphi_{+}(t)$ decays compared to its value at the beginning of
inflation $\varphi_{+}(t_{0})$. Although its exact trajectory is
initial condition dependent, one can estimate a lower bound on its
decay as long as the initial conditions have $\dot{\varphi}_{+}\lesssim-c_{+}H\varphi_{+}$
which is an attractor solution: 
\begin{equation}
\varphi_{+}(t_{k_{\mbox{min}}})=\varphi_{+}(t_{0})\left(\frac{a_{\mbox{scale}}(t_{0})}{a_{\mbox{scale}}(t_{k_{\mbox{min}}})}\right)^{\frac{3}{2}-\nu}
\end{equation}
Since $\varphi_{+}(t_{0})\lesssim M_{p}$, this sets an upper bound
on $\varphi_{+}(t_{k_{\mbox{min}}})$ for the validity of the analytic
computation:
\begin{equation}
\frac{\varphi_{+}(t_{k_{\mbox{min}}})}{M_{P}}\lesssim(\mbox{accuracy})^{\left(\frac{1}{2}-\frac{\nu}{3}\right)}.
\end{equation}

To compute the $\delta a$ correlator using theorem 3, we need the
mass of $a$. We can obtain the mass by writing the equation of motion
for $a$, neglecting the small corrections proportional to $c_{-}$:
\begin{equation}
\left(\square-\frac{\square\varphi_{+}}{\varphi_{+}}\right)a=0\label{eq:timedependentmass}
\end{equation}
\begin{equation}
\square\varphi_{+}+\frac{1}{\sqrt{g}}\partial_{\mu}\left[\frac{a^{2}}{2\varphi_{+}^{2}}\sqrt{g}\partial^{\mu}\varphi_{+}\right]-\frac{1}{2\varphi_{+}}\frac{1}{\sqrt{g}}\partial_{\mu}\left[\sqrt{g}a\partial^{\mu}a\right]+c_{+}H^{2}\varphi_{+}=0.\label{eq:approxEOM}
\end{equation}
The fact that $a$ acquires a time dependent mass is important because
that means that the decay of the $a$ field due to the mass will stop
after a finite time period: i.e. because $a$ is a NG boson, its mass
will shut off in the vacuum. Hence, we see that in the limit of $a/\varphi_{+}\ll1$,
we find
\begin{equation}
\left(\square+c_{+}H^{2}\right)a\approx0\label{eq:unperturbed}
\end{equation}
\begin{equation}
\left(\square+c_{+}H^{2}\right)\varphi_{+}\approx0.\label{eq:phipluseq}
\end{equation}
There is a remarkable symmetry in this limit because of $U(1)_{PQ}$
as explained in Eq.~(\ref{eq:constant}). The mismatch between $\delta a$
and $\delta\varphi_{+}$ coming from Eq.~(\ref{eq:timedependentmass})
is suppressed by $a/\varphi_{+}\ll1$. 

Now, once $\varphi_{+}$ rolls to the minimum of Eq.~(\ref{eq:Vphiplus}),
terms proportional to $c_{-}$ dropped in Eq.~(\ref{eq:approxEOM})
will become active allowing 
\begin{equation}
\varphi_{+}=\mbox{constant}=\sqrt{2}\sqrt{\frac{\sqrt{c_{-}}}{\sqrt{c_{+}}}F_{a}^{2}-\frac{c_{-}}{h^{2}}H^{2}}.\label{eq:globalminofphiplus}
\end{equation}
We will call the time when $\varphi_{+}$ settles down to this value
such that
\begin{equation}
\left.\frac{\square\varphi_{+}}{\varphi_{+}}\right|_{t=t_{c}}\ll H^{2}\label{eq:vanishingmass}
\end{equation}
At time $t_{c}$, $a$ will become massless because it is a NG boson.
On the other hand $\varphi_{+}$mass does not shut off even after
$\varphi_{+}$ reaches its minimum. Hence, it can be shown (similarly
as in \cite{Lemoine:2009is}) that $\delta\varphi_{+}$ keeps decreasing
while $\delta a$ freezes out. Indeed, this decay of the $\varphi_{+}$
perturbations which make it negligible is similar to the reason why
one uses a Yukawa interaction to generate observable isocurvature
perturbations in the fermionic isocurvature perturbations of \cite{Chung:2013rda}:
i.e. the non-interaction piece has a blue spectrum of $n\approx7$. 

Since the axion mass seen in Eq.~(\ref{eq:axionmassistimedependent})
is time dependent, corollary 2 is useful to compute the isocurvature
spectrum. We find
\begin{equation}
\Delta_{s}^{2}(k)=\omega_{a}^{2}\left(\frac{2^{2\nu+1}|\Gamma(\nu(\sqrt{c_{+}}H))|^{2}}{\pi}\right)\left(\frac{H(t_{k_{0}})}{2\pi a(t_{k_{0}})}\right)^{2}\left(\frac{k}{k_{0}}\right)^{3-2\nu(\sqrt{c_{+}}H)-2\epsilon_{k_{0}}+O(\epsilon_{k_{0}})g_{0}(m/H)}\label{eq:isocurvature}
\end{equation}
and the QCD axion fractional density from coherent oscillations can
be estimated for $F_{a}\ll10^{17}$ GeV as (see e.g. Eq.~(14) of
\cite{Kawasaki:2013ae})
\begin{equation}
\omega_{a}\equiv\frac{\Omega_{a}}{\Omega_{cdm}}\approx W_{a}\left(\frac{a(t_{c})}{\sqrt{2}\sqrt{\varphi_{+}^{2}(t_{c})+\varphi_{-}^{2}(t_{c})}}\right)^{2}\left(\frac{\sqrt{2}\sqrt{\varphi_{+}^{2}(t_{c})+\varphi_{-}^{2}(t_{c})}}{10^{12}\mbox{ GeV}}\right)^{n_{PT}}\label{eq:fractionalcdm}
\end{equation}
where
\begin{equation}
n_{PT}\equiv1.19\,\,\,\,\,\,\,\,\,\,\,\, W_{a}\equiv\frac{3}{2}\label{eq:params}
\end{equation}
and we have used $\Omega_{cdm}h^{2}=0.12.$ Here $k_{0}$ is fixed
fiducial wave vector and $t_{k_{0}}$ is the time when that mode leaves
the horizon. Furthermore, we can use Eq.~(\ref{eq:globalmin}) to
set
\begin{equation}
\sqrt{\varphi_{+}^{2}(t_{c})+\varphi_{-}^{2}(t_{c})}\approx F_{a}\sqrt{\frac{2}{\sqrt{c_{-}c_{+}}}(c_{-}+c_{+})}\label{eq:approxFfinal}
\end{equation}
corresponding to the minimum assuming $F_{a}\gg H$. This gives a
CDM fraction of
\begin{equation}
\omega_{a}\approx W_{a}\theta_{+}^{2}(t_{k_{0}})\left(\frac{2F_{a}\sqrt{\frac{1}{\sqrt{c_{-}c_{+}}}(c_{-}+c_{+})}}{10^{12}\mbox{ GeV}}\right)^{n_{PT}}
\end{equation}
which saturates the relic bound 
\begin{equation}
F_{a}\sim\theta_{+}^{-2/n_{PT}}(t_{k_{0}})\times10^{12}\mbox{ GeV}.
\end{equation}
Unlike in the ordinary axion scenario where $H/(2\pi F_{a})$ sets
the variance of the effective classical initial condition for $\theta_{+}(t_{k_{0}})$,
here $H/(2\pi\varphi_{+}(t_{k_{0}}))\ll H/(2\pi F_{a})$ sets the
variance.%
\footnote{Hence, even with $H\sim10^{14}$ GeV, one can have $\varphi_{+}(t_{k_{0}})\sim M_{p}$
and thus tune $\theta_{+}$ as small as $10^{-5}$ without considerations
of the variance. Of course, one can even go much smaller than the
variance with 10\% tuning as well. %
} Because the axion is a non-thermal dark matter field after the end
of inflation and because Eq.~(\ref{eq:vanishingmass}) is expected
to be continually satisfied after the end of inflation due to the
weak time dependence of Eq.~(\ref{eq:globalminofphiplus}), we expect
Eq.~(\ref{eq:isocurvature}) to be a good approximation to the final
blue isocurvature spectrum in the model of \cite{Kasuya:2009up}. 

For the validity of Eq.~(\ref{eq:isocurvature}) coming from corollary
2, the wave vector $k$ must lie in the range 
\begin{equation}
k_{\mbox{min}}<k<k_{\mbox{max}}
\end{equation}
where $k_{\mbox{min}}$ and $k_{\mbox{max}}$ have parameter dependent
constraints which we now discuss. In practice, $k_{\mbox{min}}$ and
$k_{\mbox{max}}$ should be chosen to saturate the most stringent
of the bounds listed in the corollary. Since Bunch-Davies initial
conditions must be set up for the quantum fields after inflation starts,
we must have $k_{\mbox{min}}>a(t_{b})H(t_{b})$ (where $t_{b}$ is
the beginning of inflation). Since the mode exit time $t_{k}<t_{c}$
(where $t_{c}$ is the time that the time dependent mass shifts):
\begin{equation}
k_{\mbox{min}}<k_{\mbox{max}}\lesssim k_{\mbox{min}}\times(\mbox{accuracy})^{\frac{1}{3}}\left(\frac{M_{p}}{F_{a}}\frac{1}{\sqrt{2}\left(\frac{c_{-}}{c_{+}}\right)^{1/4}}\right)^{\frac{2}{3-2\nu\left(\sqrt{c_{+}}H\right)}}\label{eq:kmax0}
\end{equation}
where we required $\varphi_{+}(t)$ to be smaller than $M_{p}$ (``accuracy''
is defined by Eq.~(\ref{eq:accuracy}) associated with the attractor
assumption). If other constraints on $k_{\mbox{max}}$ that we discuss
below allows it, $k_{\mbox{max}}$ can realistically become very close
to saturating the upper bound. One can conveniently set ``accuracy''
equal to the analytic approximation accuracy $f$ that one seeks,
although in principle, they can be set independently. One of the important
constraints related to $f$ coming from Eq.~(\ref{eq:tildeNconclusion})
with $k=k_{\mbox{max }}$ is 
\begin{equation}
k_{\mbox{max}}\ll k_{0}f^{1/\nu}\left(\frac{\varphi_{+}(t_{k_{0}})}{\sqrt{2}F_{a}\left(\frac{c_{-}}{c_{+}}\right)^{1/4}}\right)^{\frac{2}{3-2\nu\left(\sqrt{c_{+}}H\right)}}.
\end{equation}
Also, because one is using a scaling approximation about the fiducial
wave vector $k_{0}$, there is a set of inflationary model dependent
validity constraints. Given the complicated parametric dependences
of these boundaries, we will summarize below and give an explicit
example.

\begin{figure}
\begin{centering}
\includegraphics{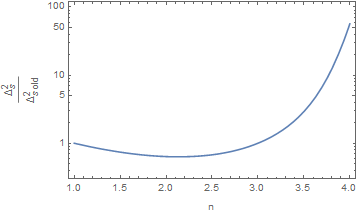}
\par\end{centering}

\protect\caption{\label{fig:Comparison-of-the}Comparison of the result of this paper
with the result of \cite{Kasuya:2009up} for the amplitude of $k=k_{0}$
(e.g. can be taken to be the CMB scale), neglecting small slow-roll
parameter corrections of $O(\epsilon)$. The enhancement is limited
by the condition of Eq.~(\ref{eq:boundonnu}). For $m/H=1.497,$one
can have $n\approx3.8$ and an enhancement of about 11.}
\end{figure}

The main correction to the original isocurvature result of \cite{Kasuya:2009up}
(i.e. Eq.~(21) after accounting for footnote 5 in \cite{Kasuya:2009up})
in the blue spectral index region (i.e. before the break) is the spectral
index dependent factor
\begin{equation}
R_{s}(c_{+})\equiv\frac{\Delta_{s}^{2}(k)}{\Delta_{s\mbox{ old}}^{2}(k)}=\frac{2^{2\nu\left(\sqrt{c_{+}}H\right)-1}|\Gamma\left(\nu\left(\sqrt{c_{+}}H\right)\right)|^{2}}{\pi}\label{eq:correctionfactor}
\end{equation}
where one must keep in mind that $\nu$ cannot be exactly zero because
of Eq.~(\ref{eq:boundonnu}).%
\footnote{Note that if one wants to compute this in a numerical setting, one
must be careful to tune the background field boundary condition $a(t_{\mbox{initial}})$
to keep the axion background field value at fiducial mode horizon
crossing $a(t_{k_{0}})$ fixed for any fixed choice of $k_{0}$ and
$m$. Also, note that as long as Eq.~(\ref{eq:boundonnu}) is obeyed,
the expansion of the Hankel function from which the $\Gamma(\nu)$
arises is a good approximation.%
} This is an interesting factor because it can give a factor of $O\left(1/(n-4)^{2}\right)$
enhancement for the blue spectrum due to the prefactor $\left|\Gamma(\nu)\right|^{2}$
while still satisfying Eq.~(\ref{eq:boundonnu}). For example, with
$m/H=1.497$, we have $n\approx3.8$, we obtain an enhancement factor
of 11. To satisfy Eq.~(\ref{eq:boundonnu}), we merely need to satisfy
\[
0.09\frac{f}{\epsilon_{k}}\gg1
\]
for inflationary scenarios with $\epsilon_{k}=10^{-4}$ even if one
desires a percent accuracy of $f=0.01$. Such enhancements are surprising
because as $\nu\rightarrow0$, one expects the quantum fluctuation
amplitude of $a$ to be smaller because its spacetime-curvature induced
tachyonic growth is smaller. Indeed, one can easily check that $\langle\delta a_{nad}\delta a_{nad}\rangle$
indeed is smaller as its effective mass controlled by $c_{+}$ is
increased. For the same parameters, the background field $a$ also
decreases as $c_{+}$ is increased. The asymptotic amplitude of $\langle\delta a_{nad}\delta a_{nad}\rangle$
associated with the Hankel function solution decreases less as $c_{+}\rightarrow3/2$
from below. This is a nontrivial difference between the quantization
induced Hankel function versus the homogeneous mode function with
slow-roll initial conditions. The comparison of the new result with
the old one is given in Fig.~\ref{fig:Comparison-of-the}. 

Next, we will summarize the explicit form of the improved spectrum
together with their validity conditions. Using Eq.~(\ref{eq:approxFfinal})
and Eq.~(\ref{eq:constant}), we can rewrite Eq.~(\ref{eq:isocurvature})
as 
\begin{eqnarray}
\Delta_{s}^{2}(k) & = & 4P_{QCD}(c_{+},c_{-},F_{a})\theta_{+}^{2}(t_{k_{0}})\times\nonumber \\
 &  & \left\{ \begin{array}{ccc}
R_{s}(c_{+})\left(\frac{H(t_{k_{0}})}{2\pi\varphi_{+}(t_{k_{0}})}\right)^{2}\left(\frac{k}{k_{0}}\right)^{3-2\nu(\sqrt{c_{+}}H)-2\epsilon_{k_{0}}+O(\epsilon_{k_{0}})g_{0}(\sqrt{c_{+}})} &  & k_{\mbox{min}}<k<k_{\mbox{max}}<k_{c}\\
\left(\frac{H(t_{k_{0}})}{2\pi\sqrt{2}F_{a}}\right)^{2}\frac{\sqrt{c_{+}c_{-}}}{c_{+}+c_{-}}\left(\frac{H(t_{0})}{H(t_{c})}\frac{k_{0}}{k_{c}}\right)^{-2\epsilon_{k_{c}}}\left(\frac{k}{k_{0}}\right)^{-2\epsilon_{k_{c}}} &  & k_{c}<k<k_{e}
\end{array}\right.\label{eq:expliciutisocurvatureresult}
\end{eqnarray}
\[
R_{s}(c_{+})\equiv\frac{2^{2\nu\left(\sqrt{c_{+}}H\right)-1}|\Gamma\left(\nu\left(\sqrt{c_{+}}H\right)\right)|^{2}}{\pi}
\]
\begin{equation}
k_{c}\equiv k_{0}\left(\frac{\varphi_{+}(t_{k_{0}})}{\sqrt{2}F_{a}\left(\frac{c_{-}}{c_{+}}\right)^{1/4}}\right)^{\frac{2}{3-2\nu(\sqrt{c_{+}}H)}}\,\,\,\,\,\,\,\,\,\,\,\,\,\,\,\,\,\,\,\,\,\,\,\,\nu(x)=\frac{3}{2}\sqrt{1-\frac{4}{9}\frac{x^{2}}{H^{2}}}
\end{equation}
\begin{eqnarray}
\frac{k_{\mbox{max}}}{k_{0}} & = & \min\left\{ f^{1/\nu}\left(\frac{\varphi_{+}(t_{k_{0}})}{\sqrt{2}F_{a}\left(\frac{c_{-}}{c_{+}}\right)^{1/4}}\right)^{\frac{2}{3-2\nu(\sqrt{c_{+}}H)}},\,\frac{k_{\mbox{min}}}{k_{0}}f^{1/3}\left(\frac{M_{p}}{F_{a}}\frac{1}{\sqrt{2}\left(\frac{c_{-}}{c_{+}}\right)^{1/4}}\right)^{\frac{2}{3-2\nu(\sqrt{c_{+}}H)}}\right.\nonumber \\
 &  & \left.,\,\exp\left[\frac{\nu^{2}(\sqrt{c_{+}}H)}{2\epsilon_{k_{0}}c_{+}}\right],\,\exp\left(\frac{f}{\epsilon_{k_{0}}}\right)\right\} 
\end{eqnarray}
\begin{equation}
\epsilon_{k_{0}}=\frac{1}{2\Delta_{\zeta}^{2}(k_{0})}\frac{1}{M_{p}^{2}}\left(\frac{H_{k_{0}}}{2\pi}\right)^{2}\label{eq:epsrelationtozeta}
\end{equation}
\begin{equation}
k_{\mbox{min}}=\max\left\{ k_{0}\exp\left(-\frac{f}{\epsilon_{k_{0}}}\right),\,\,\, a(t_{b})H(t_{b})\right\} \label{eq:otherminconstraint}
\end{equation}
\begin{equation}
P_{QCD}(c_{+},c_{-},F_{a})\equiv W_{a}^{2}2^{2n_{PT}-1}\left(\frac{c_{-}+c_{+}}{\sqrt{c_{-}c_{+}}}\right)^{n_{PT}}\left(\frac{F_{a}}{10^{12}\mbox{ GeV}}\right)^{2n_{PT}}
\end{equation}
\begin{equation}
\frac{2}{3}\sqrt{2}\pi\left(\frac{k_{H_{0}}}{k_{0}}\right)^{-\frac{3}{2}+\nu+O(\epsilon)g_{0}(m/H)}c_{+}\sqrt{\Delta_{\zeta}^{2}(k_{H_{0}})}\frac{M_{p}}{H}\frac{\theta_{+}(t_{k_{0}})}{f}<\frac{M_{p}}{\varphi_{+}(t_{k_{0}})}.\label{eq:phiplushierarchyfromquantization}
\end{equation}
where $\Delta_{\zeta}^{2}(k_{0})\approx2.4\times10^{-9}$. We have
also assumed $F_{a}\ll10^{17}$ GeV for the background axion dark
matter fraction. Some of the important background equation of motion
simplifications allowing the analytic treatment come from 
\begin{equation}
\frac{\left(c_{+}c_{-}\right)^{1/4}}{h}H\ll F_{a}\ll\varphi_{+}(t_{k_{0}})\leq\varphi_{+}(t_{k_{H_{0}}})\approx\varphi_{+}(t_{k_{0}})\left(\frac{k_{0}}{k_{H_{0}}}\right)^{\frac{3}{2}-\nu(\sqrt{c_{+}}H)}\lesssim f^{\frac{1}{2}-\frac{\nu}{3}}M_{p},\label{eq:scalehierarchies}
\end{equation}
 
\begin{equation}
\frac{H}{2\pi\sqrt{2}\varphi_{+}(t_{k_{0}})}\ll\theta_{+}(t_{k_{0}})\equiv\frac{a(t_{k_{0}})}{\sqrt{2}\varphi_{+}(t_{k_{0}})}\ll1\label{eq:thetainittuning}
\end{equation}
\begin{equation}
\frac{a(t_{c})}{\varphi_{+}(t_{c})}\ll1\longrightarrow\frac{c_{+}}{c_{-}}\ll\frac{1}{2\theta_{+}^{2}(t_{k_{0}})}-1\label{eq:cmbound}
\end{equation}
\begin{equation}
a(t_{c})<a(t_{k_{0}})\longrightarrow\sqrt{2}\frac{\sqrt{c_{-}+c_{+}}}{(c_{-}c_{+})^{1/4}}F_{a}<\varphi_{+}(t_{k_{0}})\label{eq:criticalbeinglater}
\end{equation}
where $k_{H_{0}}\approx a_{0}H_{0}\pi/2$ is the wave vector corresponding
to the observable universe today. The vector $k_{e}=a(t_{e})H(t_{e})$
is the last wave vector to leave the horizon at the end of inflation
and is typically at an unobservably small scale and is very model
dependent. This gives the improved isocurvature spectrum (together
with Eqs.~(\ref{eq:params}) and (\ref{eq:timedependentnufunc})).
The independent variables can be classified as axion-dependent parameters
$\{F_{a},\varphi_{+}(t_{k_{0}}),\theta_{+}(t_{k_{0}}),c_{\pm}\},$
inflation-dependent parameter $H(t_{k_{0}})$, and approximation scheme
dependent parameters $\{f,\, k_{0}\}$. Since physics is obviously
independent of different approximation scheme, the physical parameter
space of this model is six dimensional. The blue spectral index is
controlled by only one parameter $c_{+}$ while the break in the spectrum
is determined by $k_{\mbox{max}}$

Eq.~(\ref{eq:thetainittuning}) gives the approximate condition for
classical initial condition tuning of $\theta_{+}$ assuming there
being a quantum noise of $O(H/(2\pi))$ for the axion field because
there is no enhanced symmetry that would protect the axion field from
tadpole corrections. Although we do not address the tadpole issue
here by an explicit computation, it is reasonable to expect that the
tadpole quantum fluctuations can also significantly correct the background
equation of motion (again in the absence of enhanced symmetries) later
as $\varphi_{+}$ settles to its minimum. In that case, we should
apply a condition 
\begin{equation}
\frac{H(c_{-}c_{+})^{1/4}}{4\pi F_{a}\sqrt{c_{-}+c_{+}}}\ll\theta_{+}(t_{k_{0}})\label{eq:strongercondition}
\end{equation}
which is stronger than Eq.~(\ref{eq:thetainittuning}) because of
Eq.~(\ref{eq:criticalbeinglater}).

The isocurvature to adiabatic perturbation ratio is controlled by
\begin{eqnarray}
\frac{\Delta_{s}^{2}(k)}{\Delta_{\zeta}^{2}(k)} & = & P_{QCD}(c_{+},c_{-},F_{a})\theta_{+}^{2}(t_{k_{0}})8\epsilon_{k_{0}}\times\nonumber \\
 &  & \left\{ \begin{array}{ccc}
R_{s}(c_{+})\left(\frac{M_{p}}{\varphi_{+}(t_{k_{0}})}\right)^{2}\left(\frac{k}{k_{0}}\right)^{3-2\nu(\sqrt{c_{+}}H)+4\epsilon_{k_{0}}-2\eta_{V}} &  & k_{\mbox{min}}<k,k_{0}<k_{\mbox{max}}<k_{c}\\
\left(\frac{M_{p}}{\sqrt{2}F_{a}}\right)^{2}\frac{\sqrt{c_{+}c_{-}}}{c_{+}+c_{-}}\left(\frac{H(t_{0})}{H(t_{c})}\frac{k_{0}}{k_{c}}\right)^{-2\epsilon_{k_{c}}}\left(\frac{k}{k_{0}}\right)^{-2\eta_{V}+4\epsilon_{k_{0}}} &  & k_{c}<k<k_{e}
\end{array}\right.\label{eq:isocurvaturetocurvatureratio}
\end{eqnarray}
where the inflationary scalar spectral index is given by
\begin{equation}
n_{s}-1=2\eta_{V}-6\epsilon
\end{equation}
which can be used to phenomenologically specify $\eta_{V}$ once $n_{s}-1$
and $\epsilon$ are fixed. The CDM fraction is given by
\begin{equation}
\omega_{a}\approx W_{a}\theta_{+}^{2}(t_{k_{0}})\left(\frac{2F_{a}\sqrt{\frac{1}{\sqrt{c_{-}c_{+}}}(c_{-}+c_{+})}}{10^{12}\mbox{ GeV}}\right)^{n_{PT}}<1.\label{eq:darkmatterconstraint}
\end{equation}
Because of Eq.~(\ref{eq:generalamplitudebound}), phenomenologically
allowed parameters are in the regime of $\omega_{a}\ll1$. If one
accomplishes this with bringing down $F_{a}$, then Eq.~(\ref{eq:scalehierarchies})
brings down $H$. This in turn brings down $\varphi_{+}(t_{k_{0}})$
because of Eq.~(\ref{eq:phiplushierarchyfromquantization}).

\begin{figure}
\begin{centering}
\includegraphics[scale=0.4]{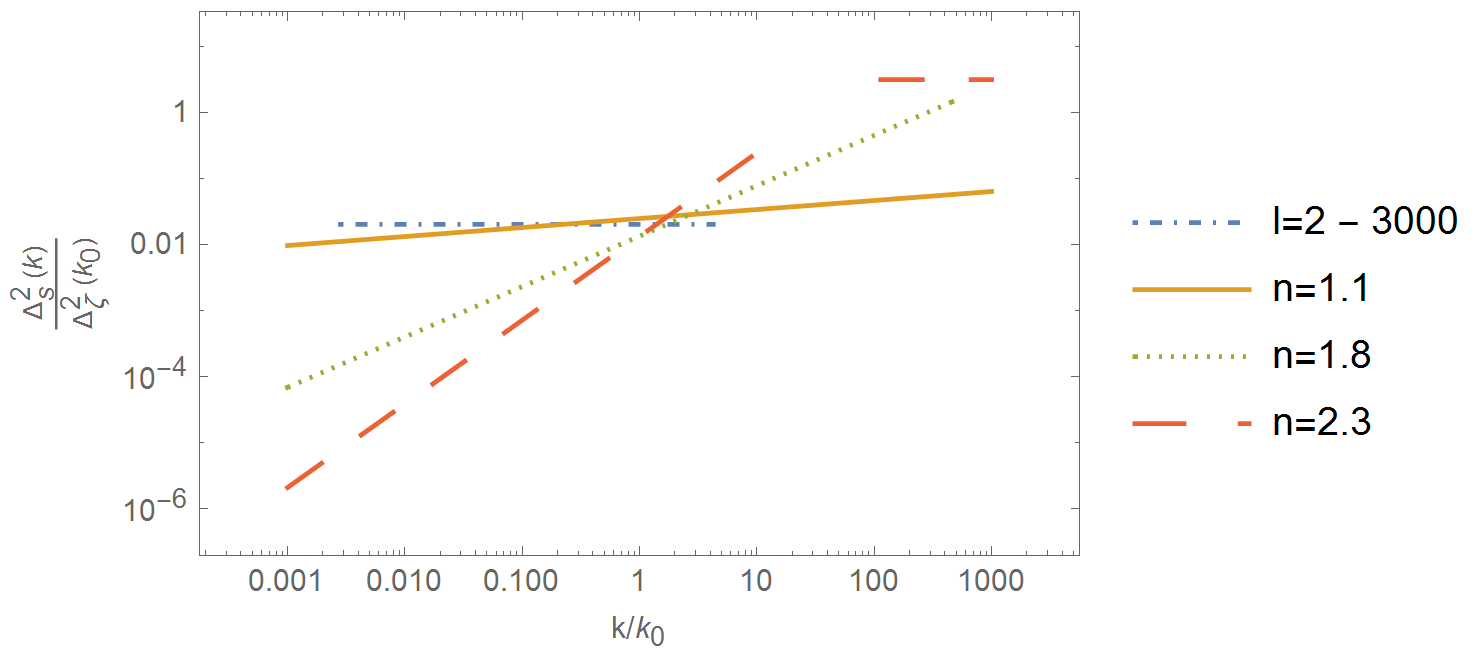}
\par\end{centering}

\protect\caption{\label{fig:spectra_func_of_cp}Illustration of the blue isocurvature
spectra for several values of the parameter $c_{+}\in\{0.2,\,1,\,1.5\}$
which map to isocurvature \{spectral index, axion dark matter fraction
$\omega_{a}$\} of $\{\{1.1,\,4.73\times10^{-4}\},\,\{1.8,\,4.06\times10^{-4}\},\,\{2.3,\,4.13\times10^{-4}\}\}$,
respectively. The rest of the parameters are fixed as given in Eq.\ (\ref{eq:Pcplusparam}),
where in particular, the dark matter fraction is. The gap in the dashed
curve in the range $k/k_{0}\in[10,100]$ occurs as a result of breakdown
of the analytic approximation associated with the fact that the effective
time-dependent mass transition occurs before the modes classicalize.
Note that the $c_{+}$ controls both the inflation induced Hubble
scale mass for the non-vacuum axion as well as the amplitude of the
spectra. That is why the three curves do not meet at a point. The
dot-dashed CMB curve represents a flat spectrum with an amplitude
of about 2\% of the adiabatic spectrum extending from $l=2-3000$
scale with $k_{0}=0.05$ Mpc$^{-1}$. Because of the transfer function
suppressing isocurvature power relative to the adiabatic power on
short length scales (an effect not shown here), the observational
constraints on these example spectra are weak.}
\end{figure}

Let us illustrate this formula with concrete parametric choices. We
plot $\Delta_{s}^{2}(k)/\Delta_{\zeta}^{2}(k_{0})$ in Fig.~\ref{fig:spectra_func_of_cp}
for $c_{+}\in\{0.2,\,1,\,1.5\}$ with the rest of the parameters fixed
at
\begin{equation}
c_{-}=0.9,\,\theta_{+}=0.04,\, F_{a}=7.9\times10^{10}\mbox{ GeV},\,\varphi_{+}(t_{k_{0}})=8.3\times10^{-7}M_{p},\, H=6\times10^{9}\mbox{ GeV},\, f=0.2.\label{eq:Pcplusparam}
\end{equation}
This set of $c_{+}$ map to \{spectral index, axion dark matter fraction
$\omega_{a}$\} of $\{\{1.1,\,4.73\times10^{-4}\},\,\{1.8,\,4.06\times10^{-4}\},\,\{2.3,\,4.13\times10^{-4}\}\}$,
respectively. To compare with the approximate CMB length scales in
the plot, we have fixed $k_{0}=0.05$ Mpc$^{-1}$.%
\footnote{Given that this paper is a paper focused on analytic computation of
the spectra, we leave more detailed numerical data fitting work to
the future.%
} The gap in the dashed curve in the range $k/k_{0}\in[10,100]$ occurs
as a result of breakdown of the analytic approximation associated
with the fact that the effective time-dependent mass transition occurs
before the modes classicalize. The actual spectrum in this gap is
not addressed by the techniques of this paper. A similar breakdown
of the assumptions leads to the termination of the dotted curve.

\begin{figure}
\begin{centering}
\includegraphics[scale=0.35]{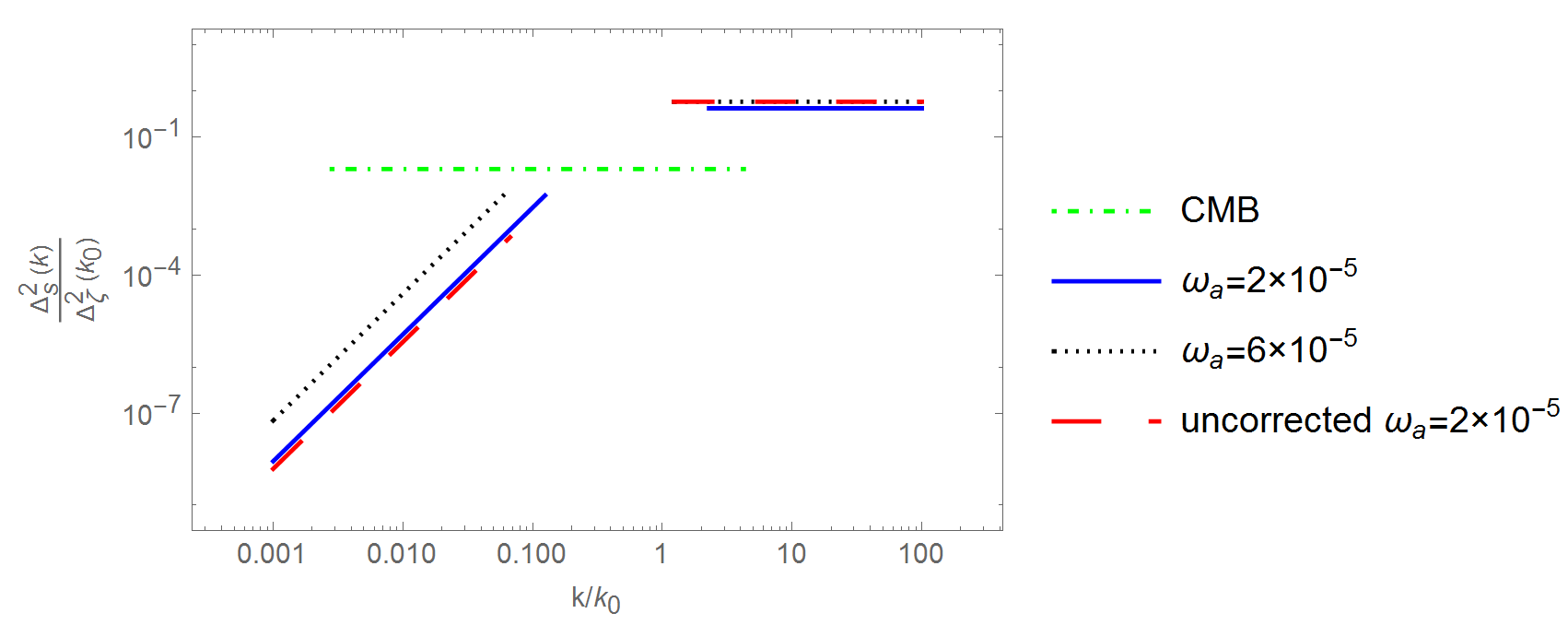}
\par\end{centering}

\protect\caption{\label{fig:spectra_func_of_Fa}Isocurvature spectra for $F_{a}=\{6.8\times10^{10}\mbox{ GeV},\mbox{ }1.6\times10^{11}\mbox{ GeV}\}$
corresponding to the axion dark matter fraction of $\omega_{a}=\{2\times10^{-5},\,6\times10^{-5}\}$.
The other parameters are fixed at values given in Eq.~(\ref{eq:PFaparam}),
and the resulting isocurvature spectral index is $n=3.8$. The ``uncorrected''
label refers to the plot that would have been given based on the previous
literature without Eq.~(\ref{eq:correctionfactor}). Note that without
the correction, one would misidentify an experimental signal of $\omega_{a}=2\times10^{-5}$
for that of $\omega_{a}=6\times10^{-5}$. The CMB label and the gaps
in the spectra are explained in Fig.~\ref{fig:spectra_func_of_cp}.
Note that parameters were chosen such as to stay close to being observable.
It is easy to choose parameters such that this spectrum is unobservable
in the foreseeable future.}
\end{figure}

Next, we illustrate what happens when the isocurvature spectral index
is very steep such that the correction factor of Eq.~(\ref{eq:correctionfactor})
becomes around a factor of 10. In Fig.~\ref{fig:spectra_func_of_Fa},
we plot the isocurvature spectra for $F_{a}=\{6.8\times10^{10}\mbox{ GeV},\mbox{ }1.6\times10^{11}\mbox{ GeV}\}$
corresponding to the axion dark matter fraction of $\omega_{a}=\{2\times10^{-5},\,6\times10^{-5}\}$,
with the other parameters fixed at 
\begin{equation}
c_{+}=2.235,\, c_{-}=0.9,\,\theta_{+}=10^{-2},\,\varphi_{+}(t_{k_{0}})=10^{-7}M_{p},\, H=9\times10^{9}\mbox{ GeV},\, f=0.7.\label{eq:PFaparam}
\end{equation}
The relatively extreme $c_{+}$ parametric choice gives an isocurvature
spectral index of $n=3.8$ and the spectral amplitude has an analytic
approximation error at the level of around 70\%.%
\footnote{A smaller choice of $f$ (corresponding to a smaller approximation
error) leads to the spectrum not being computable analytically in
the interesting region.%
} It is remarkable that dark matter fraction $\omega_{a}$ as small
as those considered in Figs.~\ref{fig:spectra_func_of_cp} and \ref{fig:spectra_func_of_Fa}
can generate potentially observable effects in cosmology. 

Another question one might have is how large $H$ can be in these
scenarios, since $H$ controls the amplitude of the tensor perturbations
that may be observable by future experiments \cite{Adam:2014bub,Flauger:2014qra,Ade:2014xna,Jinno:2014qka,Sasaki:1995aw,Starobinsky:1979ty}.
Unlike in the the ordinary axion situation, the effective PQ symmetry
breaking VEV is much larger throughout the course of the isocurvature
field evolution. This means that $H/(\mbox{time dependent PQ symmetry breaking VEV})$
which controls the effective amplitude of the isocurvature is much
smaller for observable $k$ region for the same value of $F_{a}$
which controls the axion dark matter fraction $\omega_{a}$. The parametric
tension still arises as we will now see. Combining the stabilization
constraints of $\varphi_{+}$ and $b$ (i.e. Eqs.~(\ref{eq:stabilize})
and (\ref{eq:keepatmin})) with the dark matter bound of Eq.~(\ref{eq:darkmatterconstraint})
gives
\begin{equation}
W_{a}\theta_{+}^{2}(t_{k_{0}})\left(\frac{2H/h}{10^{12}\mbox{ GeV}}\right)^{n_{PT}}\ll\omega_{a}<1.\label{eq:DMbd}
\end{equation}
Another constraint comes from the isocurvature bound of Eq.~(\ref{eq:isocurvaturetocurvatureratio}),
where we restrict to the relevant subset of constraint 
\begin{equation}
\frac{\Delta_{s}^{2}(k_{0})}{\Delta_{\zeta}^{2}(k_{0})}<\alpha_{k_{0}}\label{eq:isoeffectivebound}
\end{equation}

Both the dark matter bound of Eq.~(\ref{eq:DMbd}) and the isocurvature
bound of Eq.~(\ref{eq:isoeffectivebound}) can be satisfied for sufficiently
small $\theta_{+}(k_{0})$. To find the minimum $\theta_{+}(k_{0})$
allowed by the other constraints, note Eq.~(\ref{eq:cmbound}) (coming
from decoupling axion mixing) and Eq.~(\ref{eq:strongercondition})
(coming from the neglect of quantum tadpoles), and Eq.~(\ref{eq:thetainittuning})
imply
\begin{equation}
\frac{H^{2}}{8\pi^{2}}\frac{\sqrt{c_{-}c_{+}}}{F_{a}^{2}(c_{-}+c_{+})}\ll2\theta_{+}^{2}(t_{k_{0}})\ll\min\left\{ \frac{c_{-}}{c_{+}},\,2\right\} .\label{eq:squeezeit}
\end{equation}
Hence, there is a minimum $c_{-}$ for which this can be satisfied:
\begin{equation}
2>\frac{(c_{-})_{\mbox{min}}}{c_{+}}>\frac{1}{24}\left(2^{2/3}\left(Y(H/F_{a})\right)^{1/3}+\frac{32}{\left(\frac{1}{2}Y(H/F_{a})\right)^{1/3}}-16\right)
\end{equation}
\begin{equation}
Y(H/F_{a})\equiv\frac{27H^{4}}{\pi^{4}F_{a}^{4}}+\frac{3H^{2}\sqrt{768\pi^{4}+81H^{4}/F_{a}^{4}}}{\pi^{4}F_{a}^{2}}+128
\end{equation}
Putting this into Eq.~(\ref{eq:squeezeit}) gives 
\begin{equation}
\mbox{min}(\theta_{+}^{2})=\mu\left(\frac{F_{a}}{H}\right)\equiv\frac{1}{48}\left(2^{2/3}\left(Y(H/F_{a})\right)^{1/3}+\frac{32}{\left(\frac{1}{2}Y(H/F_{a})\right)^{1/3}}-16\right).\label{eq:minthetaplus}
\end{equation}
where the right hand side is valid whenever it is smaller than unity.
Now, the isocurvature bound in the form of Eq.~(\ref{eq:isoeffectivebound})
combined with Eq.~(\ref{eq:minthetaplus}) give
\begin{equation}
W_{a}^{2}2^{2n_{PT}-1}\left(\frac{2\mu\left(\frac{F_{a}}{H}\right)+1}{\sqrt{\mu\left(\frac{F_{a}}{H}\right)}}\right)^{n_{PT}}\left(\frac{F_{a}}{10^{12}\mbox{ GeV}}\right)^{2n_{PT}}\mu\left(\frac{F_{a}}{H}\right)8\epsilon_{k_{0}}R_{s}(c_{+})\left(\frac{M_{p}}{\varphi_{+}(t_{k_{0}})}\right)^{2}<\alpha_{k_{0}}.\label{eq:intermedbd}
\end{equation}
The $F_{a}$ dependent function on the left hand side of Eq.~(\ref{eq:intermedbd})
is a monotonically increasing function of $F_{a}$ and it is also
a monotonically increasing function of $H$. Hence, to maximize $H$,
we need to determine the smallest $F_{a}$ that is allowed by our
constraints.

One blunt constraint comes from the fact that $\mu$ is supposed to
be a small angle. Since the small angle squared $\mu(F_{a}/H)$ is
a monotonically decreasing function of $F_{a}$, one can find the
smallest $F_{a}$ allowed by the small angle assumption by solving
$\mu(F_{1}/H)=1$:
\begin{equation}
\frac{F_{1}}{H}\approx0.05.
\end{equation}
Another constraint comes from combining Eq.~(\ref{eq:DMbd}) with
Eq.~(\ref{eq:minthetaplus}). Defining $F_{2}=\min(F_{a})$ subject
to this constraint, one finds numerically
\begin{equation}
\frac{F_{2}}{H}\approx0.2\label{eq:F2def}
\end{equation}
which is larger than $F_{1}/H$. One also finds numerically that $h\gtrsim0.3$
up to the perturbative limit in this corner of allowed parametric
region. Because the $F_{a}$ dependence on $H$ bound is weak, there
is only a small shift in using $F_{2}$ versus $F_{1}$. In any case,
since $F_{2}$ is a stronger constraint, we set $F_{a}=F_{2}$ in
Eq.~(\ref{eq:intermedbd}) and need to solve for the $H$ upper bound.
Note that setting $F_{a}=F_{2}$ corresponds to a $\min\left(\theta_{+}^{2}\right)=0.04$
and $c_{-}=0.08c_{+}$.%
\footnote{Note that this does not mean that the smallest $\theta_{+}^{2}$ for
all parts of the parameter space is $0.04.$ It is only when $F_{a}$
is minimized subject to the constraints discussed, do we have this
minimum on $\theta_{+}^{2}$ .%
}

To solve for the $H$ upper bound, we still need to set $c_{+}$.
Note although $c_{+}\approx1.3795$ gives the smallest Eq.~(\ref{eq:correctionfactor})
that shows up in Eq.~(\ref{eq:intermedbd}), we must still check
the constraints coming from Eq.~(\ref{eq:scalehierarchies}). Eq.~(\ref{eq:stabilize})
part of the constraint in this parametric corner can be written explicitly
as
\begin{equation}
\frac{\left(c_{+}c_{-}\right)^{1/4}}{h}H\ll F_{a}\longrightarrow\frac{\left(2\mu(0.2)\right)^{1/4}\sqrt{c_{+}}}{h}\ll0.2\label{eq:part1const}
\end{equation}
where one sees the explicit $H$ independence because of Eq.~(\ref{eq:F2def}).
Another can be written as
\begin{equation}
\frac{\varphi_{+}(t_{k_{0}})}{M_{p}}\lesssim f^{\frac{1}{2}-\frac{\nu}{3}}\left(\frac{k_{0}}{k_{H_{0}}}\right)^{-\frac{3}{2}+\nu(\sqrt{c_{+}}H)}.\label{eq:part2const}
\end{equation}
Finally, perturbativity requires
\begin{equation}
h<\sqrt{4\pi}.\label{eq:part3const}
\end{equation}
These three constraints constrain $\{h,\, c_{+},\,\varphi_{+}(t_{k_{0}})\}$
for a fixed non-physics parameters such as $k_{0}$ and $f$.%
\footnote{The choice of $k_{0}$ does contain phenomenological information in
where the data constraints lie because one usually wants to choose
$k_{0}$ where the data is accurate.%
} Eqs.~(\ref{eq:part1const}) and (\ref{eq:part3const}) put a bound
of
\begin{equation}
c_{+}(\mbox{for max }H)<1.78\label{eq:cpluslim}
\end{equation}
corresponding to $n<2.6$. In this $c_{+}$ range, the $c_{+}$ dependence
of Eq.~(\ref{eq:intermedbd}) is weak. From Eq.~(\ref{eq:intermedbd}),
we thus conclude that the maximum $H$ that is allowed by the present
scenario is 
\begin{equation}
H\lesssim5\times10^{12}\mbox{ GeV}\left(R_{s}(c_{+})\right)^{-50/219}\left(\frac{\alpha_{k_{0}}}{3\times10^{-2}}\right)^{50/219}\left(\frac{\varphi_{+}(t_{k_{0}})/M_{p}}{10^{-1}}\right)^{100/219}.\label{eq:hubblebound}
\end{equation}

\begin{figure}
\begin{centering}
\includegraphics[scale=0.7]{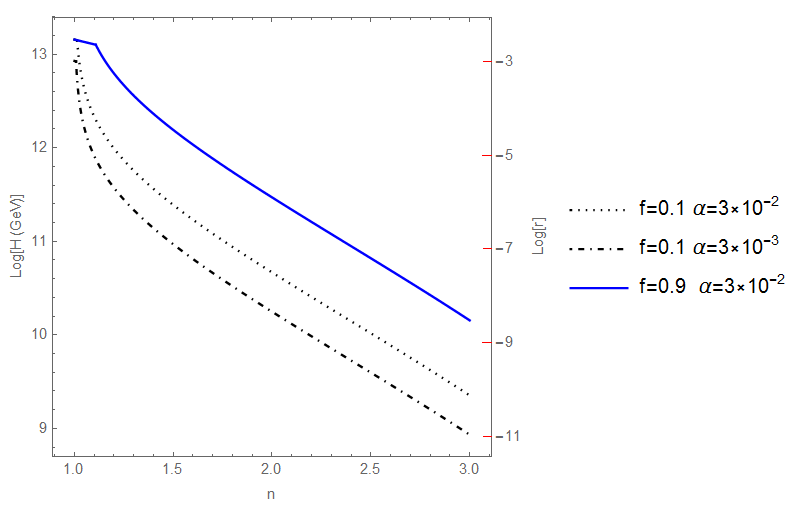}
\par\end{centering}

\protect\caption{\label{fig:Maximum--as}Maximum $H$ as a function of the isocurvature
spectral index $n$. On the right axis, the tensor-to-scalar power
spectra ratio $r$ is plotted. The break feature for small $n$ in
the $f=0.9$ case corresponds to the situation where Eq.~(\ref{eq:part2const})
becomes more important than the constraints coming from Eqs.~(\ref{eq:hubblebound})
and (\ref{eq:quantizationcond})}
\end{figure}

Next, we still need to impose the constraints of Eq.~(\ref{eq:phiplushierarchyfromquantization})
from the quantization. This constraint requires $H$ to be large and
$\varphi_{+}/M_{p}$ to be small. We can thus denote this constraint
as a lower bound on $H$:
\begin{equation}
\frac{0.82}{f}\times\left(3\times10^{-3}\right)^{-\frac{3}{2}+\nu+O(\epsilon)g_{0}(m/H)}\left(\frac{c_{+}}{0.1}\right)\left(\frac{\varphi_{+}(t_{k_{0}})/M_{p}}{0.1}\right)8.7\times10^{11}\mbox{ GeV}<H.\label{eq:quantizationcond}
\end{equation}
For any fixed $c_{+}$, both Eqs.~(\ref{eq:hubblebound}) and (\ref{eq:quantizationcond})
can be satisfied if $\varphi_{+}(t_{k_{0}})/M_{p}$ is small enough.
With the right hand side of Eq.~(\ref{eq:hubblebound}) set equal
to the left hand side of Eq.~(\ref{eq:quantizationcond}), we can
find another constraint on maximum $\varphi_{+}(t_{k_{0}})/M_{p}$
similar to that coming from Eq.~(\ref{eq:part2const}). This latter
constraint usually is more important than Eq.~(\ref{eq:part2const})
for a blue spectral index. We thus arrive at the maximum $H$ shown
in Fig.~\ref{fig:Maximum--as} for the blue spectral index scenario
considered in this section. The corresponding tensor-to-scalar power
spectra ratio $r$ is also shown on the right axis of the same figure.
It is clear that inflationary scenarios consistent with a very blue
isocurvature spectra (e.g. $n\gtrsim2.0$) will not generate tensor
spectra that is observable in the near future. Conversely, a tensor-to-scalar
ratio at the level of $r=O(10^{-1})$ will disfavor this class of
models.

Let us now summarize the how one obtained the upper bound on $H$.
One minimizes the angle $\theta_{+}(t_{k_{0}})$ consistent with the
validity of the classical equations of the motion and subject to the
decoupling constraints. Putting this into the isocurvature formula
and minimizing the isocurvature to curvature ratio by varying $F_{a}/H$
subject to dark matter abundance constraints lead to Eq.~(\ref{eq:hubblebound})
bound on $H$. The decoupling constraints are not particularly fundamental,
but the study of that region is beyond the scope of this paper. Hence,
even higher values of the tensor-to-scalar ratio may be valid with
blue isocurvature perturbations, but the phenomenological signatures
will be more complicated than the simple situation presented here.

\subsection{\label{sub:Do-Dressing-Effects}Do Dressing Effects Give a Lower
Bound on the Blue Isocurvature Spectrum?}

From the definition of $S_{\chi}$ in Eq.~(\ref{eq:isocurvaturequantity}),
we would naively expect 
\begin{equation}
\langle S_{\chi}S_{\chi}\rangle\sim4\langle\frac{\delta\chi^{(N)}}{\chi_{0}}\frac{\delta\chi^{(N)}}{\chi_{0}}\rangle+4\langle\zeta\zeta\rangle\left(\frac{\dot{\chi}_{0}}{H\chi_{0}}\right)^{2}+\mbox{cross terms}\label{eq:dressednaive}
\end{equation}
where the second term arises from the ``dressing'' effect of the
isocurvature coming from the fact that spectator isocurvature is a
contrast between the isocurvature field and the adiabatic field. For
scale invariant isocurvature spectra (such as massless axions), we
have $\dot{\chi}_{0}=0$ which means that this second term coefficient
is negligible. However, for a blue spectrum, this coefficient is of
order unity. Hence, one would naively expect 
\begin{equation}
\langle S_{\chi}S_{\chi}\rangle\gtrsim\langle\zeta\zeta\rangle\,\,\,\,\,\,\mbox{naive operator product expectation for blue spectra}\label{eq:naiveexpect}
\end{equation}
independently of $\langle\delta\chi^{(N)}\delta\chi^{(N)}\rangle/\chi_{0}^{2}$
amplitude.%
\footnote{This does not by itself give a bound on the total isocurvature perturbations
which depend on the dark matter fraction $\omega_{\chi}$ .%
} Hence, a little puzzle arises whether one can conclude
\begin{equation}
\frac{\Delta_{s_{\chi}}^{2}}{\Delta_{\zeta}^{2}}\gtrsim1
\end{equation}
for a blue isocurvature spectrum in the limit $H\rightarrow0$ with
$m/H=O(1)$ and $\chi_{0}(t_{k_{0}})$ fixed. 

Theorem 3 trivially states that this lower bound with $\chi_{0}(t_{k_{0}})$
fixed does not exist since 
\begin{equation}
\frac{\Delta_{s_{\chi}}^{2}}{\Delta_{\zeta}^{2}}\approx1.3\times10^{-2}\times2^{2\nu}\times|\Gamma(\nu)|^{2}\left(\frac{\epsilon}{10^{-2}}\right)\left(\frac{\chi_{0}}{M_{p}}\right)^{-2}
\end{equation}
which vanishes as the inflaton slow-roll parameter $\epsilon\rightarrow0$
with $\langle\zeta\zeta\rangle$ fixed. That means that 
\begin{equation}
\frac{\langle S_{\chi}S_{\chi}\rangle}{\langle\zeta\zeta\rangle}\rightarrow0
\end{equation}
for blue spectra is possible if $\chi_{0}(t_{k_{0}})$ can be fixed
contrary to the naive expectation from Eq.~(\ref{eq:naiveexpect}).%
\footnote{This limit $\epsilon\rightarrow0$ can be fraught with strong coupling
issues in the density perturbation computation formalism. We can neglect
these issues and can take this limit formally since the point is that
it decreases towards zero and not about the absolute magnitude.%
} On the other hand, consistent quantization does give a lower bound
shown in Eq.~(\ref{eq:generalamplitudebound}). However, Eq.~(\ref{eq:generalamplitudebound})
arises because $\chi_{0}(t_{k_{0}})/M_{p}$ cannot be fixed because
of the quantization approximation used: i.e. the reason for Eq.~(\ref{eq:generalamplitudebound})
is different from the naive operator multiplication analysis. Indeed,
$\Delta_{s_{\chi}}^{2}(k_{0})$ shown in Eq.~(\ref{eq:mainfinresult})
is independent of $\epsilon_{k_{0}}$ (up to the inequalities implied
by Eq.~(\ref{eq:errorestimate})) while $\Delta_{\zeta}^{2}(k_{0})$
shown in Eq.~(\ref{eq:slow-rollresult}) manifestly does depend on
$\epsilon_{k_{0}}$.

If one looks at the details of the proof to try to understand where
the naive operator multiplication analysis goes wrong, one notes that
there is a secular growth effect in which the isocurvature field $\delta\chi^{(G)}$
develops an adiabatic piece due to a secular superhorizon source effect.
For example, in the spatially flat gauge, the gravitational interaction
transmitted adiabatic piece is given by Eq.~(\ref{eq:secularpiece}).
The fact that this comes from gravitational physics can be understood
from the fact that the off-diagonal mass squared in Eq.~(\ref{eq:mixing})
are Planck suppressed, and these terms are responsible for Eq.~(\ref{eq:secularpiece}).
This can be interpreted as the effect of gravity imprinting dominant
energy inhomogeneity information onto the subdominant isocurvature
field. Hence, even if $\delta\chi^{(G)}$ at the quantum fluctuation
level does not have $\zeta$ correlation information%
\footnote{In the proof of theorem 3, the spatially flat subhorizon modes are
essentially decoupled from the inflaton modes by $\chi_{0}/M_{p}$.
This allows one to determine the Bunch-Davies state quantum correlator
independently of the sourced mixing with the inflaton in the subhorizon
region.%
}, it will on far superhorizon scales look like a mixture of adiabatic
and nonadiabatic field, in precisely the combination to eliminate
the $\zeta$ dependent pieces in $\langle S_{\chi}S_{\chi}\rangle$.

\section{Summary and Conclusion}

\label{sec:conclusion}

In this paper, we have presented three theorems and related corollaries
concerning blue spectra produced by linear spectator isocurvature
fields that give rise to CDM-photon isocurvature perturbations. Theorem
1 defines a superhorizon conserved quantity for systems possessing
an approximate symmetry of $V_{\chi}'(\delta\chi)\approx V_{\chi}''(\delta\chi)\delta\chi$.
The merit of this theorem compared to previous discussions of this
topic in the literature is its ability to go beyond the end of inflation
and the reheating process. Theorem 2 describes under what averaging
conditions that fluid quantities behave as $\delta\chi_{nad}/\chi_{0}$.
This second theorem merely restates what is known in the literature
(see e.g.~\cite{Gordon:2000hv,Polarski:1994rz}) in the context of
current theorems. Theorem 3 describes the computation of the quantum
isocurvature perturbations. The merit of theorem 3 compared to the
previous discussion in the literature is the explicit canonical quantization
in the presence of linearized gravitational constraints. The validity
regime of this theorem imposes a nontrivial constraint of Eq.~(\ref{eq:nontrivial}).
If this condition is violated, the amplitude of quantization is expected
to be more complicated than the simple analytic treatment presented
here.

In Sec.~\ref{sub:Improvement-of-0904.3800}, we have applied the
theorems to the work of \cite{Kasuya:2009up} and improved their computation.
In the process, we have uncovered a conserved Noether current associated
with $U(1)_{PQ}$ that is leading to the tracking of the axion field
with the radial direction field. The final spectral formula is given
in Eq.~(\ref{eq:expliciutisocurvatureresult}). The general magnitude
comparison of the isocurvature blue part of the spectrum is shown
in Fig.~\ref{fig:Comparison-of-the}. The spectral break features
and the validity of analytic computations were explored in Figs.~\ref{fig:spectra_func_of_cp}
and \ref{fig:spectra_func_of_Fa}. The maximum tensor-to-scalar ratio
for which this simple scenario remains valid is shown in Fig.~\ref{fig:Maximum--as}.

In Sec.~\ref{sub:Do-Dressing-Effects}, we have applied the theorems
to explain how naive operator product estimates for the isocurvature
correlator amplitude lower bound fails. The main physics is that the
spectator field attains the inhomogeneities associated with the inflaton
through its gravitational coupling. From a perturbation theory perspective,
this inhomogeneity is attained through a secular effect which would
naively be dropped from the consideration of perturbative expansion
coefficient alone. From a physical perspective, the spectator field
which undergoes no appreciable quantum fluctuations by themselves
still attains an inhomogeneity that looks like the inflaton's inhomogeneities.
Interestingly, we do uncover in this paper an isocurvature correlator
amplitude lower bound Eq.~(\ref{eq:generalamplitudebound}) whose
phenomenological validity requires the dark matter fraction $\omega_{\chi}$
to be much smaller than unity. If this lower bound is violated, the
quantization of the isocurvature perturbations do not take on the
simple form presented in this paper.

There are many possible future extensions of this work. Regarding
the general applicability of the theorem, it would be interesting
to find interaction strength boundaries for classes of models for
which the linear spectator behavior of the isocurvature perturbation
survives. Regarding the scenario of \cite{Kasuya:2009up}, we have
explicitly laid out where the analytic computation fails near the
break region of the spectra. Although we would naively expect that
either side of the break region to be smoothly connected, we would
also naively expect features to exist in that region. Numerical investigations
of the features may be interesting for discovery potential. Other
obvious future investigation possibilities include improving our understanding
of the experimental discovery prospects of the blue spectral isocurvature
perturbations.
\begin{acknowledgments}
This work was supported in part by the DOE through grant DE-FG02-95ER40896.
This work was supported in part by the Kavli Institute for Cosmological
Physics at the University of Chicago through grant NSF PHY-1125897
and an endowment from the Kavli Foundation and its founder Fred Kavli.
\end{acknowledgments}

\appendix

\section{\label{sec:Particular-Solution-In}Particular Solution In the Subhorizon
Region}

Consider the approximate equation of motion object for $\delta\chi^{(sf)}$
coming from Eq.~(\ref{eq:matrixeom})
\begin{equation}
E\equiv\frac{d^{2}\delta\chi^{(sf)}}{dt^{2}}+3H\partial_{t}\delta\chi^{(sf)}+\left(\frac{k^{2}}{a^{2}}+M_{22}^{2}\right)\delta\chi^{(sf)}+m^{2}\frac{\dot{\varphi}_{0}\chi_{0}}{M_{p}^{2}H}\delta\varphi^{(sf)}\label{eq:EOMoperator}
\end{equation}
where we have approximated 
\begin{equation}
M_{21}^{2}=m^{2}\frac{\dot{\varphi}_{0}\chi_{0}}{M_{p}^{2}H}\left[1+O\left(\frac{\chi_{0}}{M_{P}}\right)\right].
\end{equation}
\emph{Without} taking the superhorizon limit, consider the following
particular solution ansatz:
\begin{equation}
\delta\chi^{(sf)}=\frac{\dot{\chi}_{0}}{\dot{\varphi}_{0}}\delta\varphi^{(sf)}.\label{eq:particularansatz}
\end{equation}
The equation of motion object $E$ becomes
\begin{equation}
E=\frac{\dot{\chi}_{0}}{\dot{\varphi_{0}}}\left(\partial_{t}^{2}\delta\varphi+\left[3H-2m^{2}\frac{\chi_{0}}{\dot{\chi}_{0}}+2\frac{V_{\varphi}'(\varphi)}{\dot{\varphi}_{0}}\right]\delta\dot{\varphi}+W\delta\varphi\right)
\end{equation}
\begin{equation}
W\equiv\frac{k^{2}}{a^{2}}+M_{22}^{2}-m^{2}-6m^{2}\frac{\chi_{0}}{\dot{\chi}_{0}}H+\frac{m^{2}}{M_{p}^{2}}\frac{\chi_{0}\dot{\varphi}_{0}^{2}}{H\dot{\chi}_{0}}+6H\frac{V_{\varphi}'(\varphi_{0})}{\dot{\varphi}_{0}}-2\frac{m^{2}\chi_{0}V_{\varphi}'(\varphi_{0})}{\dot{\chi}_{0}\dot{\varphi}_{0}}+2\frac{[V_{\varphi}'(\varphi_{0})]^{2}}{\dot{\varphi}_{0}^{2}}+V_{\varphi}''(\varphi_{0}).
\end{equation}
Next, the usual slow-roll approximation gives
\begin{equation}
2\frac{V_{\varphi}'(\varphi)}{\dot{\varphi}_{0}}\approx-6H
\end{equation}
\begin{equation}
-2m^{2}\frac{\chi_{0}}{\dot{\chi}_{0}}\approx6H\label{eq:anotherslow-roll}
\end{equation}
which implies
\begin{equation}
W\approx\frac{k^{2}}{a^{2}}+M_{22}^{2}-m^{2}+(3\eta_{V}-6\epsilon)H^{2}.
\end{equation}
Note that Eq.~(\ref{eq:anotherslow-roll}) assumes a small slow-roll
factor analogous to $\epsilon$ but for the $\chi_{0}$ field. We
will call these expansion parameters
\begin{equation}
\eta_{\chi}\equiv\frac{m^{2}}{9H^{2}}\ll1\,\,\,\,\,\,\,\epsilon_{\chi}\equiv\frac{1}{54}\frac{m^{4}}{H^{4}}\frac{\chi_{0}^{2}}{M_{p}^{2}}\ll1\label{eq:slowrollofchi}
\end{equation}
whose motivation is detailed in Sec.~\ref{sec:Slow-Roll-ofchi}.
Explicitly, one can show 
\begin{equation}
-2m^{2}\frac{\chi_{0}}{\dot{\chi}_{0}}+2\frac{V_{\varphi}'(\varphi)}{\dot{\varphi}_{0}}=\left[O(\epsilon^{3/2})+O(\sqrt{\epsilon}\eta_{V})+O(\sqrt{\epsilon}\eta_{\chi})\right]H
\end{equation}
 Next. we know
\begin{equation}
M_{22}^{2}=m^{2}[1+O\left(\frac{\chi_{0}^{2}}{M_{P}^{2}}\right)]
\end{equation}
 which yields
\begin{equation}
W\approx\frac{k^{2}}{a^{2}}+(3\eta_{V}-6\epsilon)H^{2}.
\end{equation}
Finally, we also write down the equation of motion for $\delta\varphi^{(sf)}$
as
\begin{equation}
\frac{d^{2}\delta\varphi^{(sf)}}{dt^{2}}+3H\partial_{t}\delta\varphi^{(sf)}+\left(\frac{k^{2}}{a^{2}}+(3\eta_{V}-6\epsilon)H^{2}\right)\delta\varphi^{(sf)}+m^{2}\frac{\dot{\varphi}_{0}\chi_{0}}{M_{p}^{2}H}\delta\chi^{(sf)}=0
\end{equation}
and note that 
\begin{equation}
m^{2}\frac{\dot{\varphi}_{0}\chi_{0}}{M_{p}^{2}H}\delta\chi^{(sf)}=m^{2}\mbox{sgn}\dot{\varphi}_{0}\frac{\sqrt{2\epsilon}\chi_{0}}{M_{p}}\delta\chi^{(sf)}=O\left(\frac{\chi_{0}}{M_{p}}\right)\delta\chi^{(sf)}.
\end{equation}
We thus conclude
\begin{equation}
E=0+O\left(\frac{\chi_{0}}{M_{p}}\right)+O\left(\epsilon^{n>1}\right)+O(\eta_{\chi}\sqrt{\epsilon})\label{eq:ordercount}
\end{equation}
which means that Eq.~(\ref{eq:particularansatz}) solves the equation
of motion even in the subhorizon region in the leading approximation. 

Finally, note that the particular solution itself is of order 
\begin{equation}
\frac{\dot{\chi}_{0}}{\dot{\varphi}_{0}}\delta\varphi^{(sf)}=O\left(\frac{\chi_{0}}{M_{p}}\right)\delta\varphi^{(sf)}
\end{equation}
which means that since we are dropping the same order in Eq.~(\ref{eq:ordercount}),
one might naively think there is no content in the solution. However,
note that on the left hand side of Eq.~(\ref{eq:EOMoperator}), there
is a factor of $(k/a)^{2}$ which makes this term non-negligible.
The point of this section was to show that such unsuppressed terms
are all canceled by the slow-roll equations of motion to leading order
in slow-roll expansion.

\section{\label{sec:Mixture}Mixture}

Result of theorem 3 is applicable only when the dark matter $\chi$
can be made to be totality of dark matter. Suppose
\begin{equation}
\frac{\delta\rho_{cdm}}{\rho_{cdm}}=\frac{\delta\rho_{X}+\delta\rho_{Y}}{\rho_{X}+\rho_{Y}}
\end{equation}
where $X$ can be the $\chi$ particle and $Y$ is the rest of the
cold dark matter. We can rewrite this as
\begin{eqnarray}
\frac{\delta\rho_{cdm}}{\rho_{cdm}} & = & \frac{\rho_{X}\delta_{X}+\rho_{Y}\delta_{Y}}{\rho_{X}+\rho_{Y}}\\
 & = & \omega_{X}\delta_{X}+\omega_{Y}\delta_{Y}.
\end{eqnarray}
where
\begin{equation}
\omega_{X}\equiv\frac{\rho_{X}}{\rho_{X}+\rho_{Y}}\,\,\,\,\,\,\,\,\,\,\,\omega_{Y}\equiv\frac{\rho_{Y}}{\rho_{X}+\rho_{Y}}\label{eq:omegachi}
\end{equation}
such that $\omega_{X}+\omega_{Y}=1$. Hence
\begin{eqnarray}
\delta_{S} & = & \frac{\delta\rho_{cdm}}{\rho_{cdm}}-\frac{3}{4}\frac{\delta\rho_{\gamma}}{\rho_{\gamma}}\\
 & = & \omega_{X}\delta_{X}+\omega_{Y}\delta_{Y}-\frac{3}{4}\delta_{\gamma}
\end{eqnarray}
If
\begin{equation}
\delta_{Y}=\frac{3}{4}\delta_{\gamma},
\end{equation}
then
\begin{equation}
\delta_{S}=\omega_{X}\left[\delta_{X}-\frac{3}{4}\delta_{\gamma}\right]=\omega_{X}\delta_{S_{X}}\label{eq:effectofmixing}
\end{equation}
where $\delta_{S_{X}}$ is the isocurvature in $X$ component. Hence,
with mixing, the isocurvature is diluted by a factor $\omega_{X}$.

\section{\label{sec:Slow-Roll-ofchi}Slow-Roll of $\chi$}

In this section, we motivate a slow-roll expansion parameter for $\chi$
field. Consider
\begin{equation}
\ddot{\chi_{0}}+3H\dot{\chi}_{0}+V_{\chi}'(\chi_{0})=0\label{eq:secondordereom}
\end{equation}
where $H$ is an externally determined time dependent function. Divide
through by $HM_{p}^{2}$.
\begin{equation}
\frac{\ddot{\chi}_{0}}{HM_{p}^{2}}+3\frac{\dot{\chi}_{0}}{M_{p}^{2}}+\frac{V_{\chi}'(\chi_{0})}{HM_{p}^{2}}=0\label{eq:scaledsecondordereom}
\end{equation}

Define a formal hierarchy
\begin{equation}
\frac{\ddot{\chi}_{0}}{HM_{p}^{2}}\ll\left\{ 3\frac{\dot{\chi}_{0}}{M_{p}^{2}},\,\frac{V_{\chi}'(\chi_{0})}{HM_{p}^{2}}\right\} 
\end{equation}
 and expand $\dot{\chi}_{0}/M_{p}^{2}$ about zero using a formal
perturbation parameter $\lambda$:
\begin{equation}
\{\frac{\dot{\chi}_{0}}{M_{p}^{2}}=O(\lambda),\,\,\,\,\frac{V_{\chi}'(\chi_{0})}{HM_{p}^{2}}=O(\lambda)\}\label{eq:potentialderivativeorder}
\end{equation}
Using this expansion, construct a trial solution 
\begin{equation}
\dot{\chi}_{0}=-\frac{V_{\chi}'(\chi_{0}(t))}{3H(t)}\lambda+v_{1}\lambda^{2}.\label{eq:trial}
\end{equation}

Put the trial solution Eq.~(\ref{eq:trial}) into Eq.~(\ref{eq:secondordereom}).
For this endeavor, we need to evaluate $\ddot{\chi}_{0}$ to second
order in $\lambda$:
\begin{eqnarray}
\ddot{\chi}_{0} & = & -\frac{V_{\chi}''(\chi_{0}(t))}{3H(t)}\dot{\chi}_{0}\lambda+\frac{V_{\chi}'(\chi_{0}(t))}{3H^{2}}\dot{H}\lambda+\lambda^{2}\dot{v}_{1}\\
 & = & \frac{V_{\chi}''(\chi_{0}(t))}{3H(t)}\frac{V_{\chi}'(\chi_{0}(t))}{3H(t)}\lambda^{2}+\frac{V_{\chi}'(\chi_{0}(t))}{3H^{2}}\dot{H}\lambda+\lambda^{2}\dot{v}_{1}+O(\lambda^{3})
\end{eqnarray}
To finish expanding the right hand side, we need to evaluate $\dot{H}$.
According to the usual slow-roll, we have
\begin{equation}
\dot{H}=-\epsilon H^{2}.
\end{equation}
 Hence, we conclude
\begin{equation}
\ddot{\chi}_{0}=\frac{1}{9H^{2}}V_{\chi}'(\chi_{0})V_{\chi}''(\chi_{0})\lambda^{2}-\lambda\epsilon\frac{V_{\chi}'(\chi_{0})}{3}+\lambda^{2}\dot{v}_{1}+O(\lambda^{3})
\end{equation}

Put the trial solution Eq.~(\ref{eq:trial}) into Eq.~(\ref{eq:scaledsecondordereom}),
account for the second term of Eq.~(\ref{eq:potentialderivativeorder}),
treat $\epsilon\lambda=O(\lambda^{2})$ in the formal counting, and
collect $O(\lambda)$ and $O(\lambda^{2})$:
\begin{equation}
O(\lambda^{2}):\,\,\,\frac{1}{9H^{2}}V_{\chi}'(\chi_{0})V_{\chi}''(\chi_{0})-\frac{\left[V_{\chi}'(\chi_{0})\right]^{3}}{54H^{4}M_{p}^{2}}+\dot{v}_{1}+3\frac{\sqrt{V_{\chi}}}{\sqrt{3}M_{p}}v_{1}=0.
\end{equation}
This can be solved
\begin{equation}
v_{1}=e^{-3\int dtH}\int dt\frac{\left[V_{\chi}'(\chi_{0})\right]^{3}}{54H^{4}M_{p}^{2}}e^{3\int dtH}-e^{-3\int dtH}\int dt\frac{1}{9H^{2}}V_{\chi}'(\chi_{0})V_{\chi}''(\chi_{0})e^{3\int dtH}.
\end{equation}
Hence, we conclude the fractional correction to $\dot{\phi}$ is
\begin{equation}
\frac{v_{1}}{\frac{V_{\chi}'(\chi_{0}(t))}{3H(t)}}=O\left(\frac{1}{\frac{V_{\chi}'(\chi_{0}(t))}{3H(t)}}\times\frac{1}{3H}\times\frac{1}{9H^{2}}V_{\chi}'(\chi_{0})V_{\chi}''(\chi_{0})\right)+O\left(\frac{1}{\frac{V_{\chi}'(\chi_{0}(t))}{3H(t)}}\times\frac{1}{3H}\times\frac{\left[V_{\chi}'(\chi_{0})\right]^{3}}{54H^{4}M_{p}^{2}}\right).
\end{equation}
This motivates us to define the following slow-roll parameters: 
\begin{eqnarray}
\frac{1}{\frac{V_{\chi}'(\chi_{0})}{3H}}\times\frac{1}{3H}\times\frac{1}{9H^{2}}V_{\chi}'(\chi_{0})V_{\chi}''(\chi_{0}) & = & \frac{1}{9H^{2}}V_{\chi}''(\chi_{0})\equiv\eta_{\chi}\\
\frac{1}{\frac{V_{\chi}'(\chi_{0}(t))}{3H(t)}}\times\frac{1}{3H}\times\frac{\left[V_{\chi}'(\chi_{0})\right]^{3}}{54H^{4}M_{p}^{2}} & = & \frac{\left[V_{\chi}'(\chi_{0})\right]^{2}}{54H^{4}M_{p}^{2}}\equiv\epsilon_{\chi}.
\end{eqnarray}
Obviously, this is not unique, and other slow-roll definitions exist.
See e.g. \cite{Byrnes:2006fr}.

\bibliographystyle{JHEP}
\bibliography{ref,misc,ConsistencyRelation,Curvaton,Non-Gaussianity,Inflation_general,wimpzilla,deltaN,gravitino_isocurvature,blue_isocurvature}
 
\end{document}